\documentclass[10pt,conference]{IEEEtran}

\IEEEoverridecommandlockouts

\usepackage[normalem]{ulem}
\usepackage{amsmath}
\usepackage[english]{babel}
\usepackage{blindtext}
\usepackage{booktabs} 
\usepackage{array}
\usepackage{multirow}
\usepackage{caption}
\usepackage[FIGBOTCAP]{subfigure}
\usepackage{adjustbox}
\usepackage{enumitem}
\usepackage{xcolor}
\usepackage{balance}

\usepackage{epsfig}
\usepackage[linesnumbered,ruled,vlined]{algorithm2e}
\SetKw{Continue}{continue}
\SetKwComment{Comment}{$\triangleright$\ }{}
\SetKwInOut{Parameter}{Parameters}

\newcommand{\ie}{{\em i.e., }}
\newcommand{\eg}{{\em e.g., }}
\newcommand{\myverb}{\fontsize{10}{48}\usefont{OT1}{lmtt}{b}{n}\noindent }
\newcommand{\myverbSmall}{\fontsize{8}{48}\usefont{OT1}{lmtt}{b}{n}\noindent }
\usepackage{tcolorbox}

\newcommand{\remove}[1]{} 

\AtBeginDocument{%
  \providecommand\BibTeX{{%
    \normalfont B\kern-0.5em{\scshape i\kern-0.25em b}\kern-0.8em\TeX}}}

\ifCLASSINFOpdf
\else
\fi

\begin{document}


\title{Modeling Live Video Streaming: Real-Time Classification, QoE Inference, and Field Evaluation}	


\author{Sharat~Chandra~Madanapalli,
	Alex~Mathai, 
	Hassan~Habibi~Gharakheili,  
	and~Vijay~Sivaraman
	\IEEEcompsocitemizethanks{
		\IEEEcompsocthanksitem S. C. Madanapalli, H. Habibi Gharakheili, and V. Sivaraman are with the School of Electrical Engineering and Telecommunications, University of New South Wales, Sydney, NSW 2052, Australia (e-mails: sharat.madanapalli@unsw.edu.au, h.habibi@unsw.edu.au, vijay@unsw.edu.au).
		\IEEEcompsocthanksitem A. Mathai is with BITS Pilani (e-mail: f2016339p@alumni.bits-pilani.ac.in).
		\IEEEcompsocthanksitem This submission is an extended and improved version of our paper presented at the IEEE/ACM IWQoS 2021 conference \cite{IWQoS21}.		
	}
}

%
%
%
%
%


\IEEEtitleabstractindextext{%
\begin{abstract}
Social media, professional sports, and video games are driving rapid growth in live video streaming, on platforms such as Twitch and YouTube Live. Live streaming experience is very susceptible to short-time-scale network congestion since client playback buffers are often no more than a few seconds.  
Unfortunately, identifying such streams and measuring their QoE for network management is challenging, since content providers largely use the same delivery infrastructure for live and video-on-demand (VoD)  streaming, and packet inspection techniques (including SNI/DNS query monitoring) cannot always distinguish between the two. 

In this paper, we design, build, and deploy \textit{ReCLive}: a machine learning method for live video detection and QoE measurement based on network-level behavioral characteristics. Our contributions are four-fold: (1) We analyze about 23,000 video streams from Twitch and YouTube, and identify key features in their traffic profile that differentiate live and on-demand streaming. We release our traffic traces as open data to the public; (2) We develop an LSTM-based binary classifier model that distinguishes live from on-demand streams in real-time with over 95\% accuracy across providers; 
(3) We develop a method that estimates QoE metrics of live streaming flows in terms of resolution and buffer stall events with overall accuracies of 93\% and 90\%, respectively; and (4) Finally, we prototype our solution, train it in the lab, and deploy it in a live ISP network serving more than 7,000 subscribers.
Measurements from the field show that 99.8\% of Twitch videos are streamed live, while this measure is only 2.3\% for YouTube. Further, during peak hours as many as 15\% of live video streams are played at low-definition resolution and about 7\% of them experience a buffer stall. Our method provides ISPs with fine-grained visibility into live video streams, enabling them to measure and improve user experience.

\end{abstract}

}

\maketitle

\IEEEdisplaynontitleabstractindextext

\IEEEpeerreviewmaketitle
\vspace{2mm}

\section{Introduction} \label{sec:intro}

Live video streaming consumption grew by 65\% from 2017 to 2018 \cite{Conviva2018} and the recent pandemic situation fueled further growth with major events being streamed online \cite{livestream}. 
YouTube since 2017 allows the larger public to do live streaming, and is widely used for concerts, sporting events, and video games. Twitch (acquired by Amazon)  
is a popular platform for streaming video games from individual gamers as well as from tournaments. 
\remove{Viewership of eSport tournaments already outnumbers traditional sport tournaments like SuperBowl \cite{cnbc2019} - this citation was criticized in previous submission}Therefore, Internet Service Providers (ISPs) are keen on gaining fine-grained visibility into live video streams, enabling them to monitor quality of experience (QoE) for live video streaming over their networks, and where necessary enhance QoE for their subscribers by dimensioning bandwidth appropriately or applying policies for traffic management.

Ensuring good QoE for live video streams is challenging, since clients per-force have small playback buffers (a few seconds at most) to maintain a low latency as the content is being consumed while it is being produced. 
Even short time-scale network congestion can cause buffer underflow leading to a video stall, causing user frustration. Indeed, consumer tolerance is much lower for live than for on-demand video \cite{IABsurvey2018}, since they may be paying specifically to watch that event as it happens, and might additionally be missing the moments of climax that their social circle is enjoying and commenting on. Our partner ISP on this project has corroborated with anecdotal evidence that live-streaming consumers tend to make more support calls regarding poor experience.

Network operators lack the tools today to distinguish live streaming flows in their network, let alone know the QoE associated with them. Content providers like YouTube use the same delivery infrastructure for live streaming as for on-demand video, making it difficult for deep packet inspection (DPI) techniques to distinguish between them. Indeed, most commercial DPI appliances use DNS queries and/or SNI (Server Name Indication) certificates to classify traffic streams, but these turn out to be the same for live and on-demand video (\eg in Youtube), making them indistinguishable. In this work, we therefore pursue an alternative approach that is based on the behavioral profile of the traffic flows. Extracting key attributes from the network behavior allows us to build machine learning models that can distinguish live from on-demand video, as well as estimate user QoE metrics in terms of resolution and buffer stall events. This in turn allows network operators to detect and measure live video streams purely from network behavior, without requiring any assistance from end-clients or server/CDN end-points. 

Our specific contributions are four-fold: \textbf{First}, we collect traffic traces from field (university campus) and lab (using synthetic network conditions) of around 23,000 video streams spanning the two popular providers, namely Twitch and YouTube, and analyze their patterns (\S\ref{sec:characteristics}). 
We	make our dataset (video playback metrics and network measurements)  openly available to the research community. We make a key observation that 
requests for manifest files and media segments display markedly different patterns. Focusing on the time series of content requests, therefore, lets us develop our \textbf{second} contribution wherein an LSTM (long short term memory) model is trained to distinguish live from on-demand video with an accuracy over 95\% across both providers (\S\ref{classification}). Our \textbf{third} contribution develops a method that uses the chunk-based features collected from the network flows to estimate QoE metrics for live streaming in terms of resolution using a random forest classifier with 93\% accuracy, and predict buffer stalls using a statistical model with an accuracy of 90\% (\S\ref{sec:qoe}). For our \textbf{final} contribution, we integrate our methods into a complete system and deploy it in an ISP network serving over 7000 customers (\S\ref{sec:prototype}). By analyzing the field measurements, we augment our models with a state machine to reduce mis-classifications. 
Our system provides ISPs with real-time visibility into live video streaming and its QoE metrics.

\section{Related Work} \label{sec:prior}

\textbf{Live Video Streaming:}
Several aspects of live video streaming have been studied by researchers including QoE modeling/measurement. 
Prior work on QoE of live videos range from theoretical models \cite{9212876} to study buffer dynamics to analysis of HTTP logs of CDNs to predict resolution and buffer stalls \cite{ahmed2017suffering,guarnieri2017characterizing}.
Our work, instead, focuses on identification of video streams and predicting QoE of encrypted live streams from real-time network traffic behavior. We note that HTTP logs and QoE metrics are typically available to the CDNs or content providers. Our work is positioned to support ISPs who do not have access to such logs but require to infer the QoE of live videos traversing their network.

\textbf{Application Classification:}
Traffic classification has been a widely studied cross-disciplinary field and more recently, researchers have begun the use of machine learning/deep learning models for classification of network traffic \cite{pacheco2018towards}. Authors of \cite{iTeleScope19} classify video streaming vs. large downloads by using manually extracted features from network flow activity to train random forest classifiers. In contrast, deep learning-based methods leverage the automatic feature extraction: work in \cite{8004872} classifies type of traffic (Mail, VoIP, Chat etc.) using a CNN on the first $784$ bytes of a session and similarly work in \cite{rezaei2018achieve} classifies Google applications using CNN-based models on packet lengths and inter-arrival times. Work in \cite{8171733,8026581} attempts to detect intrusion attacks and classify IoT applications respectively by implementing neural networks combining LSTMs/RNNs \cite{10.1162/neco.1997.9.8.1735} and CNNs. The prior works developed various types of models to learn network traffic patterns at bytes or packets level. 
In our work, we use LSTM based model for classifying live and VoD streams. Unlike existing methods that use packet/byte-level features, we rely on periodic flow-level request counters (collected every half-second) 
making it relatively scalable to high traffic rates. 

\textbf{Video QoE From Network:} 
Recently, many researchers \cite{mangla2018emimic,gutterman2019requet,bronzino2019inferring,madanapalli2019inferring, 9212897} have studied QoE metrics for video streaming services across providers such as YouTube, Netflix, Facebook, Bilibili and Amazon, particularly focusing on VoD. Among existing works, only \cite{bronzino2019inferring} studied QoE for live streaming services (Twitch) by estimating only the resolution metric. We note that not only does live video differ in delivery, but it also has more stringent QoE requirements. Further, prior works predominantly performed post-facto analysis of video streaming QoE using features extracted from pcap traces \cite{mangla2018emimic, gutterman2019requet, 9212897}, or CDN logs \cite{ahmed2017suffering, guarnieri2017characterizing}.  While authors of \cite{bronzino2019inferring} evaluated their online methods via deployment in home networks (by embedding their software into home gateways), our work distinguishes itself in deployment and operational scenario. We build a network monitoring system which detects live video streaming applications and measures their QoE metrics for an ISP managing tens of gigabits-per-second of Internet traffic. In terms of QoE, our work complements and builds upon existing literature  by developing a method to detect buffer stalls by tracking the buffer health of live video streams in real-time. The QoE metrics obtained from our system can be further used to augment: routing optimization systems like \cite{9079927}, or an adaptive scheduling systems like \cite{10.1145/3341216.3342215}.
Our design choices primarily aim at scalability and ease of deployment by identifying inexpensive traffic attributes (to compute), and building machine learning models that are ``general'' (work across providers) and ``simple''  (lower-memory footprint, and ease of training and deployment).


\section{Live Video Characteristics \& Dataset} \label{sec:characteristics}


Live video streaming refers to video content which is simultaneously recorded and broadcasted in real-time. The content uploaded by the streamer sequentially passes through ingestion, transcoding, and a delivery service of a content provider before reaching the viewers (\cite{TwitchEng,pires2015youtube,zhang2015crowdsourced}). 
Modern live streaming clients typically use protocols (e.g. HTTP Live Streaming) wherein they fetch \textit{manifest} files containing URLs to the latest  media segments that are transcoded into multiple resolutions. Using adaptive bitrate algorithms, a video client then picks and streams segments of an appropriate resolution while maintaining a short buffer to keep the latency between streamer and viewer to a minimum. 
This increases the chances of buffer underflow as network conditions vary, making live videos more prone to QoE impairments such as resolution drop and video stall (\cite{ahmed2017suffering, guarnieri2017characterizing}).

In contrast, VoD streaming uses HTTP Adaptive Streaming (HAS) and involves the client requesting segments from a server which contains pre-encoded video resolutions. This not only enables the use of sophisticated multi-pass encoding schemes which compress segments to smaller sizes, but also lets the client maintain a larger buffer making it less prone to QoE deterioration.


\subsection{Network Activity Analysis}\label{sec:netanalysis}
\begin{figure*}[t!]
	\begin{center}
		\mbox{
			\subfigure[Twitch Live.]{
				{\includegraphics[width=0.29\textwidth,height=0.21\textwidth]{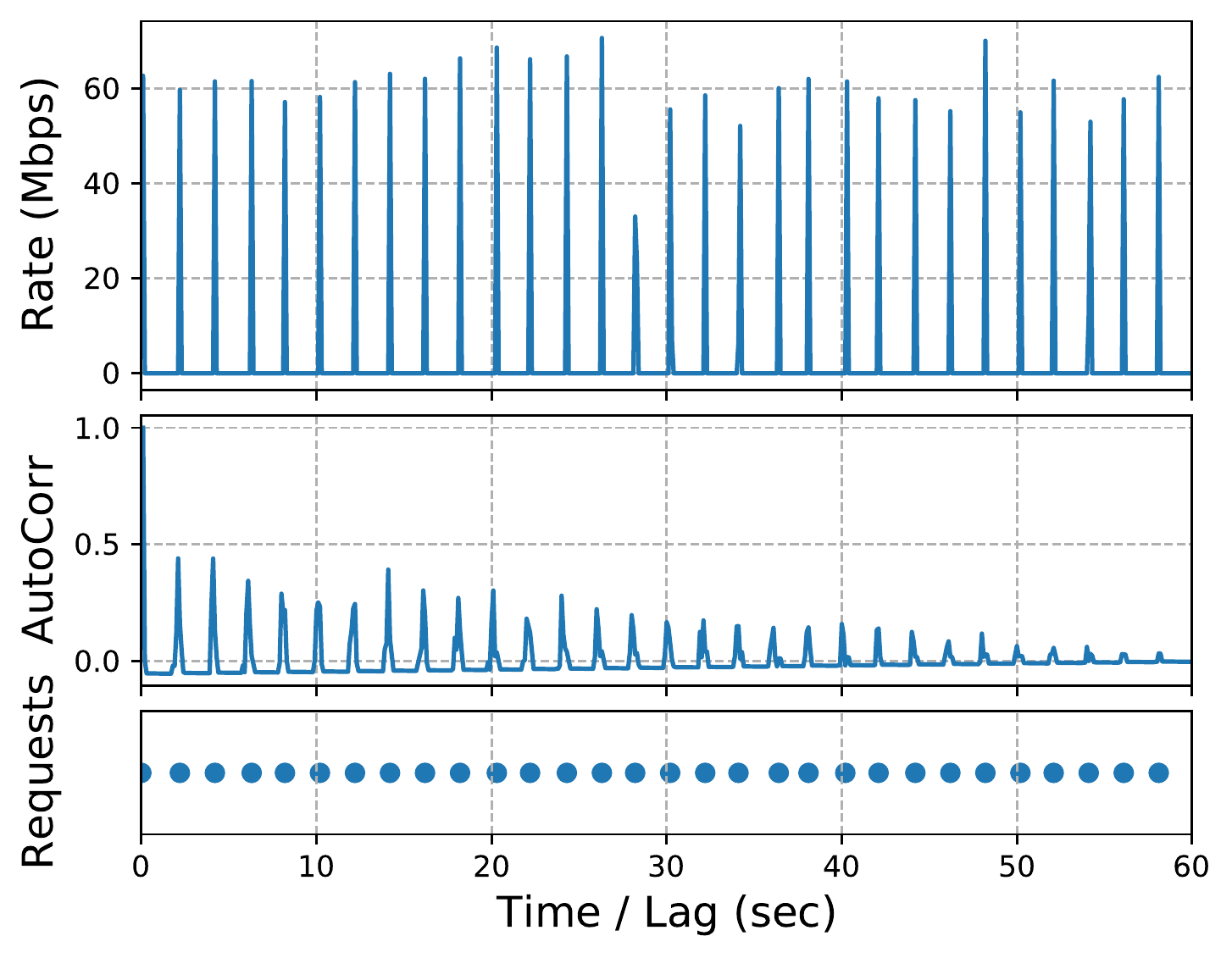}}\quad
				\label{fig:timeseriesTwitchLive}
			}
		}
		\hspace{-7mm}  
		\mbox{
			\subfigure[Twitch VoD.]{
				{\includegraphics[width=0.29\textwidth, height=0.21\textwidth]{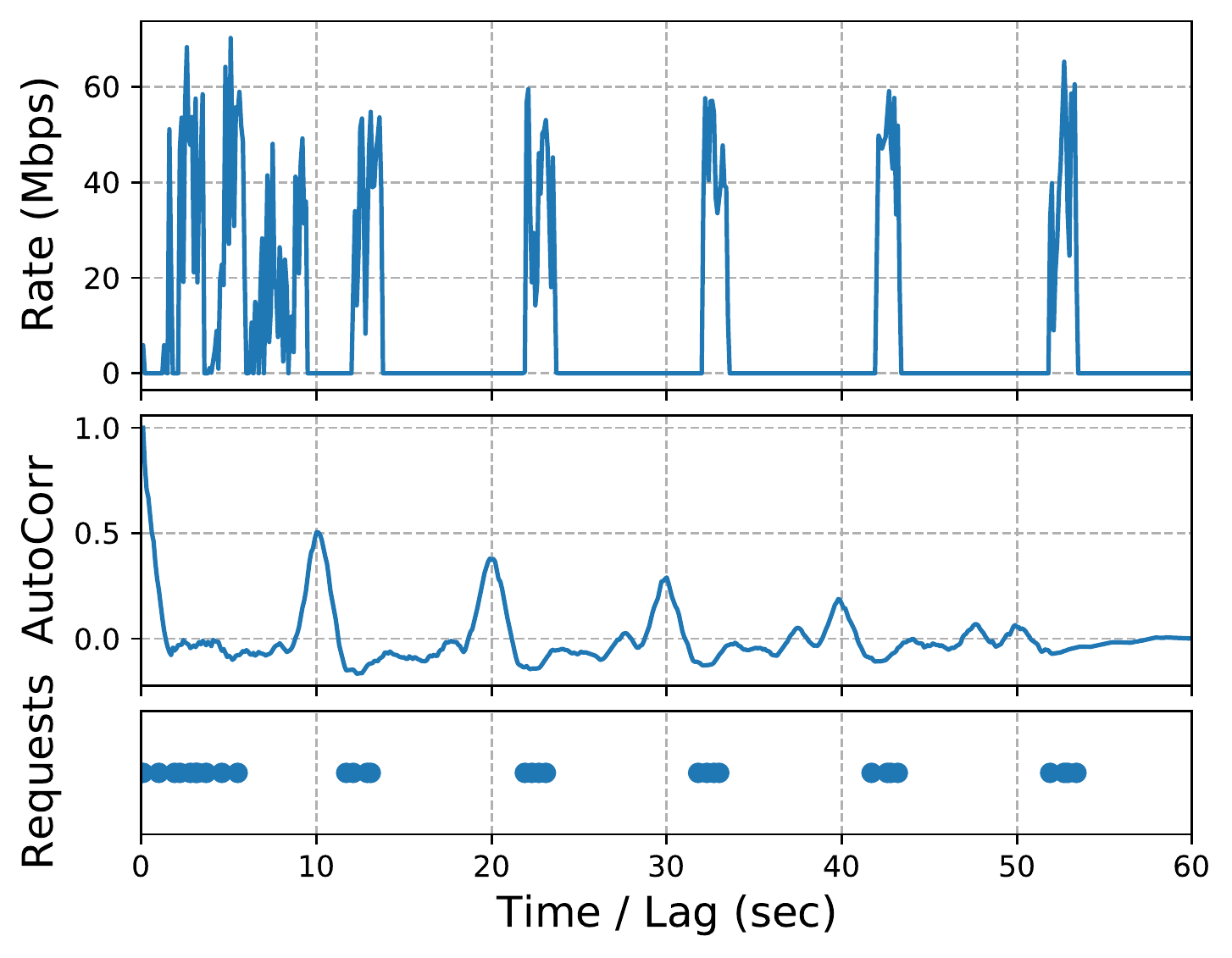}}\quad
				\label{fig:timeseriesTwitchVoD}
			}
		}
		\hspace{-7mm}  
		\mbox{
			\subfigure[YouTube Live.]{
				{\includegraphics[width=0.29\textwidth,height=0.21\textwidth]{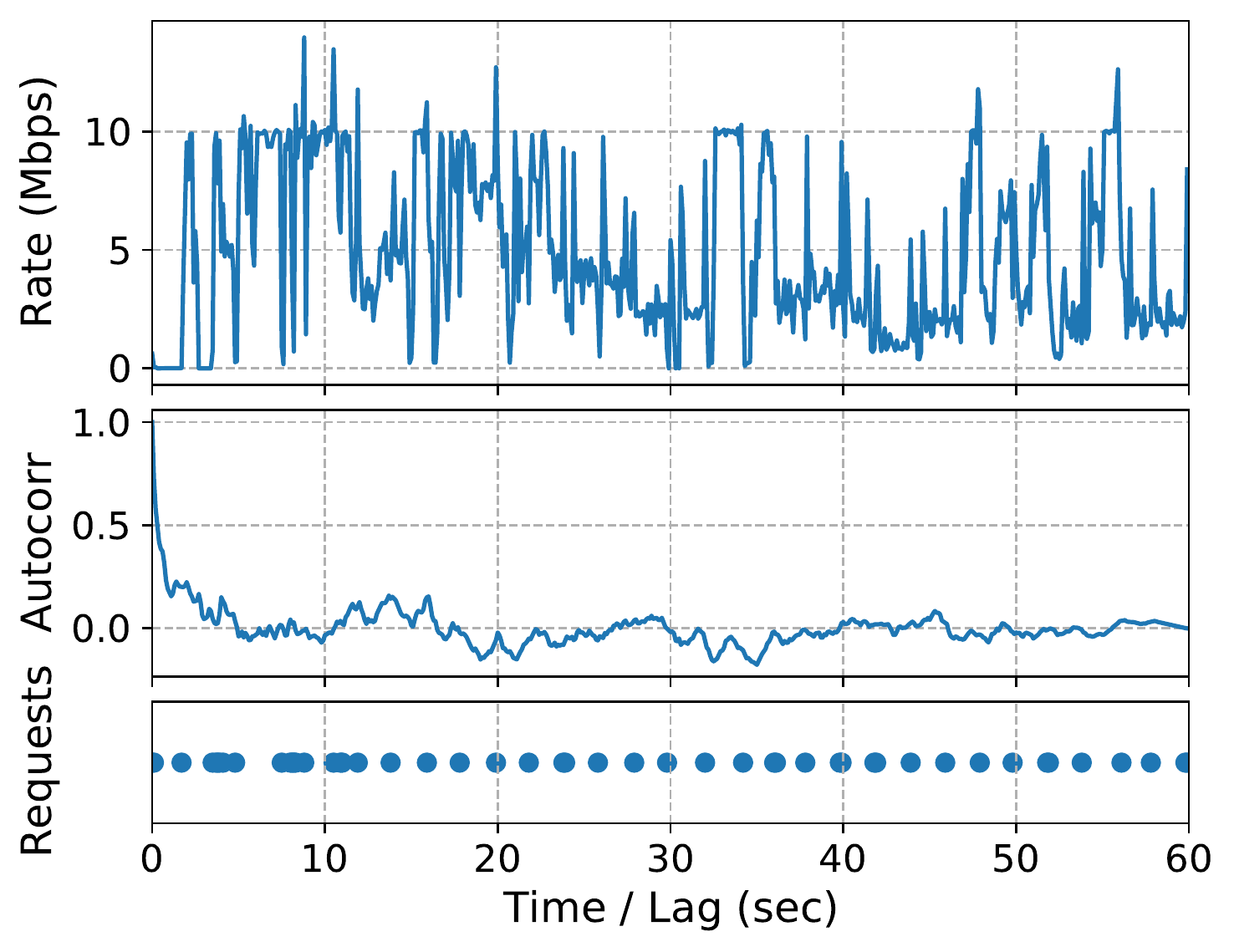}}\quad
				\label{fig:youtubelive}
			}
		}
		\hspace{-2mm}
		\vspace{-3mm}
		\caption{Network behavior: download rate profile, its auto-correlation, and request packets.}
		\vspace{-5mm}
		\label{fig:timeseries}
	\end{center}
\end{figure*}

Fig.~\ref{fig:timeseriesTwitchLive} and \ref{fig:timeseriesTwitchVoD} show the client's network behavior (download rate collected at 100 ms granularity) of representative live and VoD Twitch streams from our dataset. 
The  live streaming client downloads video segments every two seconds. In contrast, the VoD client begins by downloading multiple segments to fill up a long buffer and then fetches subsequent segments every ten seconds. Thus, the periodicity of segment downloads seems to be a very important feature to distinguish live from VoD streams.

\begin{table*}[t]
	\centering
	\caption{Fetch mechanisms of Twitch and YouTube video streaming.}
	\vspace{-1mm}
	\label{tab:vidfetch}
	\begin{tabular}{lllllll} 
		\toprule
		{Provider}         & Type & Protocol & Request for Manifest & Frequency & Latency modes & Service Endpoint SNI \\
		\midrule
		\multirow{2}{*}{Twitch}   & VoD        & HTTP/2  & Once & 10s & - &  {\myverbSmall{vod-secure.twitch.com}} \\
		& Live       & HTTP/1.1    & Periodic (different flow) & 2/4s &  Low, Normal  & {\myverbSmall{video-edge*.abs.hls.ttv.net}}\\
		\multirow{2}{*}{YouTube}  & VoD        & HTTP/2 + QUIC  & Once & 5-10s &   - & {\myverbSmall{*.googlevideo.com}}\\
		& Live       & HTTP/2 + QUIC    & Manifestless & 1/2/5s &  Ultra Low, Low  & {\myverbSmall{*.googlevideo.com}}\\
		\bottomrule 
	\end{tabular}
	\vspace{-5mm}
\end{table*}

We first estimate the periodicity for download signals by applying auto-correlation function followed by peak detection. Fig.~\ref{fig:timeseries} shows the auto-correlation value at different time lags (integral multiple of a second) for both live and VoD Twitch streams. Note that the auto-correlation sequence displays periodic characteristics just the same as the signal itself, \ie lag = 2s for live Twitch and lag = 10s for VoD Twitch.
Further, we also notice peaks at multiples of the periodicity value. From this observation, we attempt to classify video streams as live or VoD using a Random Forest classifier whose inputs are the first three lag values at which the auto-correlation signal peaks, and achieved an accuracy of about $89.5$\% for Twitch videos. However, this method does not generalize well to other content providers due to several identifiable challenges. Varying network conditions causes the auto-correlation to fail in identifying the periodicity, as shown in Fig.~\ref{fig:youtubelive} for a sample of YouTube live streaming.
Further, user triggered activities like trick-play for VoD seem to distort the time-trace signal, causing it to be mis-classified as a live stream.  


\begin{figure*}[t!]
	\begin{center}
		\mbox{
			\subfigure[Architecture of our automated data collector tool.]{
				{\includegraphics[width=0.90\columnwidth]{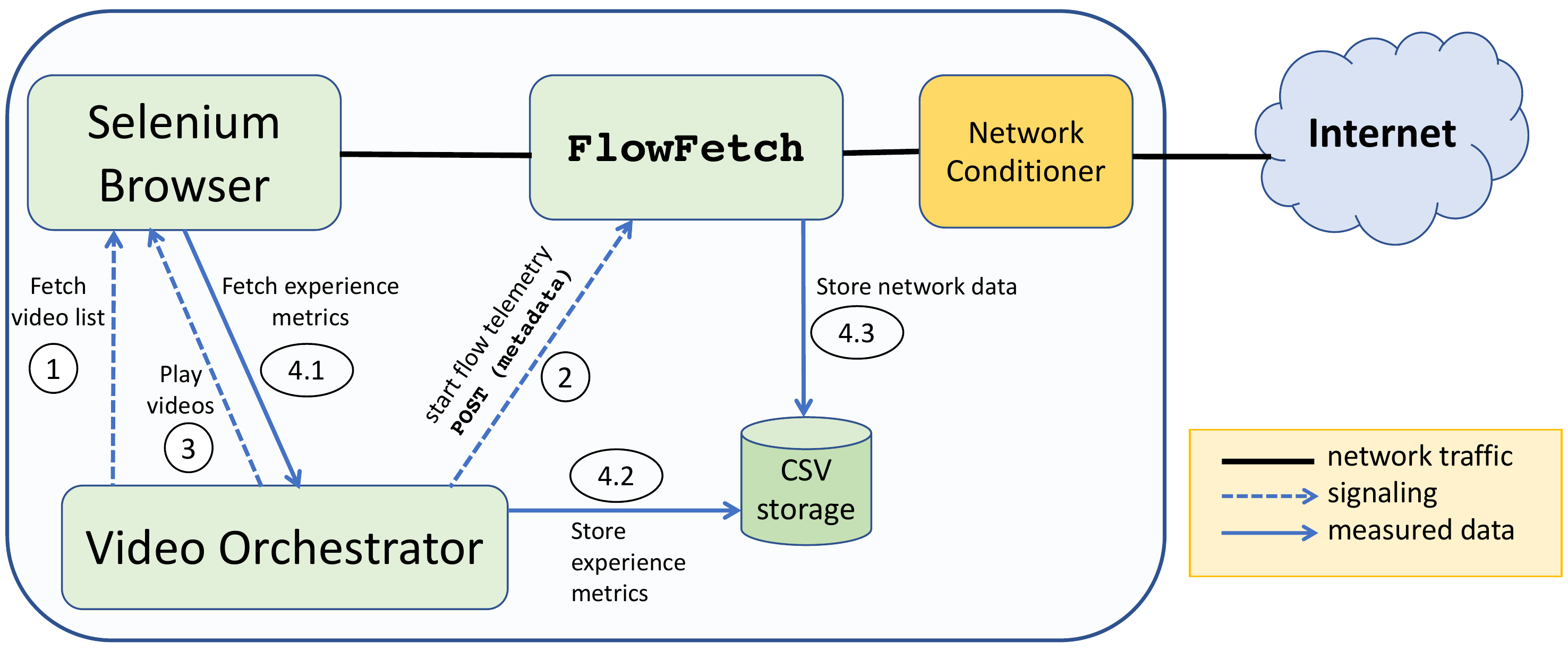}}\quad
				\label{fig:datatool}
			}
		}
		\hspace{-1mm}
		\mbox{
			\subfigure[Data collection setup in our campus network.]{
				{\includegraphics[width=0.90\columnwidth]{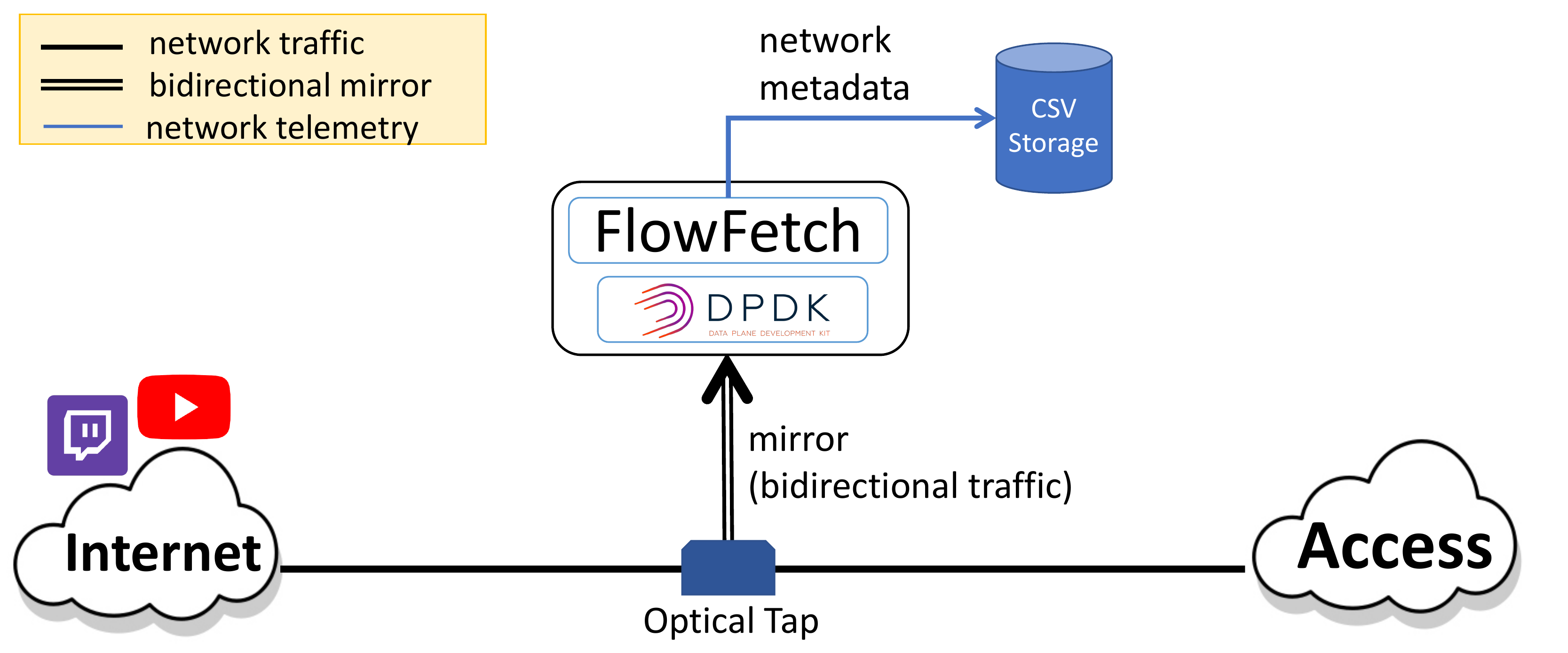}}\quad
				\label{fig:campusarch}
			}
		}
		\caption{Data collection: (a) Architecture of our tool and (b) Campus setup.}
		\vspace{-5mm}
		\label{fig:tool}
	\end{center}
\end{figure*} 

To overcome these challenges and better understand the delivery mechanism of live videos, we collected the playback data from the video client such as latency modes, buffer sizes and resolutions (using the browser automation tool described in \S\ref{sec:dataset}). We also used the network debugging tools available in Google Chrome browser and Wireshark (configured to decrypt SSL) to gain insights into protocols being used, patterns of the requests made for content and manifest files, their periodicity, and the available latency modes as shown in Table~\ref{tab:vidfetch}. Before we explain the observations for each provider, we would like to clarify that a (network) \textit{flow} is identified by the 5-tuple:  \textit{SrcIP}, \textit{DstIP}, \textit{SrcPort}, \textit{DstPort}, and \textit{Protocol}.

\textbf{Twitch}: The Twitch VoD client uses HTTP/2 and fetches combined audio and video media segments (with extension {\myverb{.ts}}) from a server endpoint with the SNI ``{\myverb{vod-secure.twitch.com}}''. 
Twitch live, however, uses HTTP/1.1, and on the same TCP flow fetches separate audio and video segments from a server with the SNI ``{\myverb{video-edge*.abs.hls.ttv.net}}''. The ``{\myverb{video-edge*}}'' part indicates the identity of edge computes of the CDN\footnote{Content Distribution Network} used by Twitch. Video providers employ CDN services to distribute transcoded segments of their contents, placing them closer to end-users to reduce the latency. In addition to media segments, the client requests manifest files from a different server with a prefix name ``{\myverb{video-weaver}}'', which also seems to be served by the CDN.
The periodicity of segment requests is around 10 seconds for VoD and 2 seconds for live streams \cite{TwitchEng}, corroborating our earlier observation in this section. We also observed a small fraction of streams in which video segments are requested at a periodicity of 4 seconds -- such cases are accounted for when predicting video QoE metrics. Additionally, Twitch offers two modes of latency, \ie Low and Normal. We note that these modes differ in: (a) client buffer capacity, which is higher (\ie 6-8 seconds) for Normal when compared to Low (\ie 2-4 seconds), and (b) the use of CMAF media containers which is is predominant in Low but rarely used in Normal.

\textbf{YouTube}: YouTube primarily uses HTTP/2 over QUIC \cite{langley2017quic} for both VoD and live streams, fetching audio and video segments separately over multiple flows (usually two in case of QUIC). These flows are established to the server endpoint with name matching pattern ``{\myverb{*.googlevideo.com}}". If QUIC protocol is disabled or not supported by the browser (\eg Firefox, Edge, or Safari), YouTube uses multiple TCP flows over HTTP/1.1 to fetch the video content. YouTube live operates in \textit{manifestless} mode (as indicated by the client playback statistics), and thus manifest files are not transferred on the network. In the case of VoD, after filling up the initial buffer, the client typically tops it up at a periodicity of 5-10 seconds. We observed that the buffer size and periodicity could vary depending on the selected resolution and/or network conditions. In the case of live streaming, however, the buffer health and the periodicity of content fetch will depend on the latency mode of the video. For YouTube live, there are three latency modes, including Ultra Low (buffer health: 2-5 sec, periodicity: 1 sec, uses CMAF media containers), Low (buffer health: 8-12 sec, periodicity: 2 sec), and Normal (buffer health: 30 sec, periodicity: 5 sec). We found that live streaming in the normal latency mode displays the same network behavior as VoD, and hence is excluded from our study -- this mode of streaming is not as sensitive as the other two modes.

As mentioned above, while SNIs may seem sufficient to distinguish live and VoD streams, at least for Twitch - this strategy does not work for YouTube. SNIs can also be changed by content providers at any time without prior notice. Further, with the increasing adoption of eSNI (encrypted SNI) supported by TLS 1.3, server names will not be accessible from the network traffic. Thus, it is necessary to identify certain patterns in live streaming applications' network behavior to distinguish them from VoD streams. We observed that for each media segment/manifest file, the video client in the browser made an HTTP request that was also seen as an upstream packet on the wire. Due to the use of TLS, the HTTP request is hidden in the packet data. Thus, we tag the upstream packets to be request packets when they contain a payload greater than 26 bytes (the minimum size of HTTP payload). Fig.~\ref{fig:timeseries} clearly illustrates how the request packets correlate with the video segments being fetched -- however, the auto-correlation approach failed to capture it (as indicated in the YouTube instance). We found that the time-trace signal of request packets: (a) is periodic and indicative of the streaming type, even in varying network conditions, (b) is less prone to noise in case of user-triggered activities, and (c) can be well generalized across content providers.

\begin{table}[t!]
	\centering
	\caption{Summary of our dataset: number of streams.}
	\vspace{-2mm}
	\label{table:dataset}
	\begin{tabular}{ccccc}
		\toprule
		\textbf{Source} & \multicolumn{2}{c}{\textbf{Twitch}} & \multicolumn{2}{c}{\textbf{YouTube}} \\
		& Live         & VoD         & Live          & VoD        \\
		\midrule
		Tool              & 2587            & 2696          &1430              &1719 \\
		Campus            & 12534            & 1948           & -             & -           \\
		\bottomrule
	\end{tabular}
	\vspace{-6mm}
\end{table}

\begin{figure*}[h]
	\begin{center}
		\mbox{
			\subfigure[An LSTM cell.]{
				{\includegraphics[width=0.65\columnwidth,height=0.4\columnwidth]{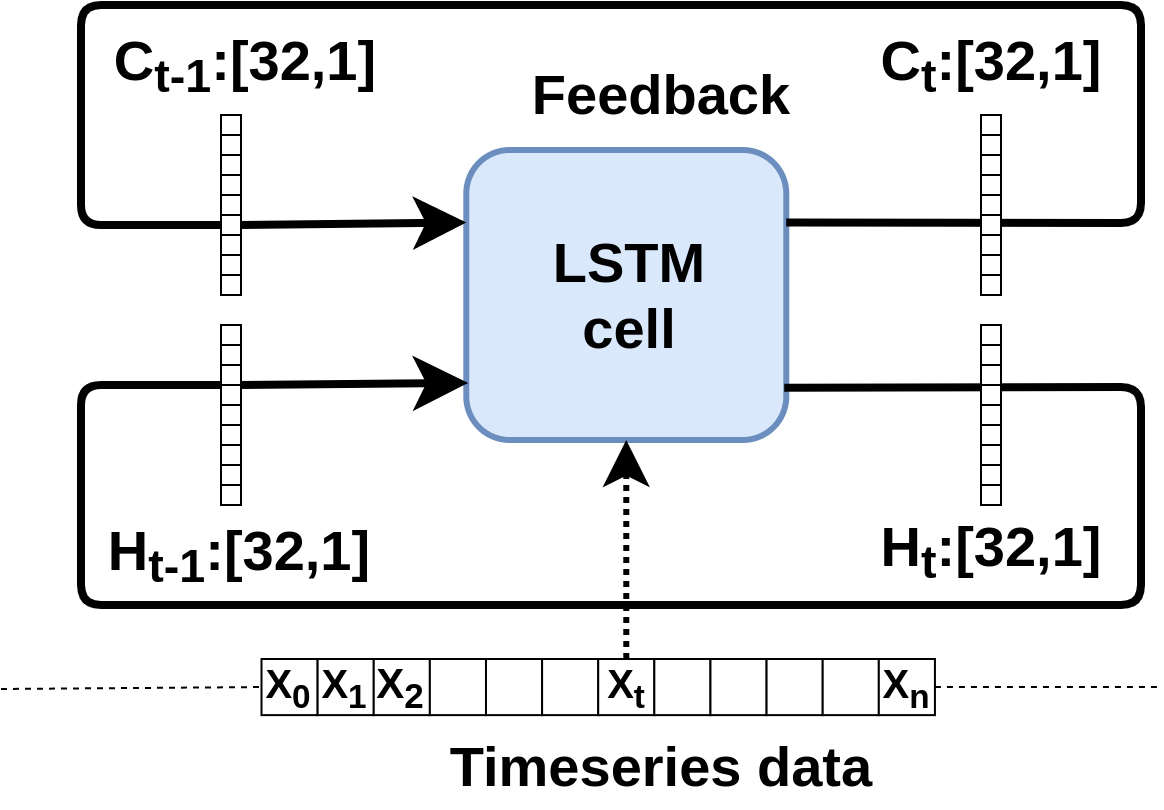}}\quad
				\label{fig:lstmcell}
			}
		}
		\mbox{
			\subfigure[LSTM to MLP network.]{
				{\includegraphics[width=0.7\columnwidth,height=0.4\columnwidth]{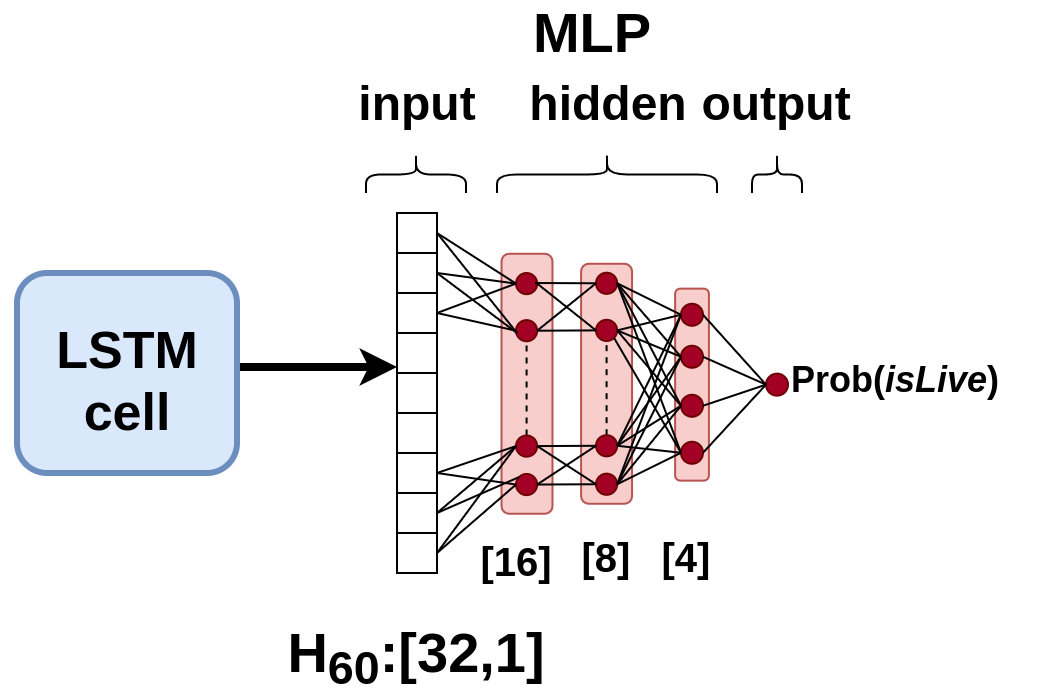}}\quad
				\label{fig:neuralnet}
			}
		}
		\vspace{-1mm}
		\caption{$Model$ structure for binary classification.}
		\vspace{-6mm}
		\label{fig:model}
	\end{center}
\end{figure*}

\vspace{-1mm}
\subsection{Dataset}\label{sec:dataset}

Having identified request packets as a key feature to distinguish live from VoD streams, we collect data of around $23,000$ video streams from Twitch and YouTube. In this section, we describe two tools which we built to: (a) automate the playback of video streams, and  (b) collect data of video streams from our campus network -- we obtained appropriate ethics clearances (UNSW Human Research Ethics Advisory Panel approval number HC16712) for this study. 
To collect necessary data for developing our models, we built a tool that records both network telemetry (flow-level counters) and user experience metrics of video streams (shown in Fig. \ref{fig:datatool}). The tool automatically plays live and VoD streams from Twitch and YouTube on personal computers. The tool has three main containerized components, including: (i) a browser controlled using the Selenium \cite{selenium} library, (ii) a network telemetry component called ``FlowFetch'', and (iii) an orchestrator that co-ordinates the playback of videos and data collection.
This tool also has a network conditioner module to artificially impose network conditions with the help of the {\myverb{tc}} tool available in Linux distributions. 

\textbf{Automated Data Collection}: The orchestrator starts by fetching a list of videos to be played. 
For both live and VoD, the tool fetches the top trending videos from a particular provider. It then iterates through the video list and performs the following steps for each video: (1) signals the FlowFetch component to start collecting network data, (2) plays the video on the browser, (3) collects the experience metrics reported by the player such as resolution, buffer level, and stores them in a csv file. During the playback of the video, FlowFetch collects and stores network data into multiple csv files. Once the video is played for a fixed duration of time ($5$ minutes in this study), the orchestrator signals FlowFetch to stop collecting data and thereafter repeats the steps on the next video.

We have developed the FlowFetch component in Golang to collect network telemetry data. It can read packets from a pcap file, a network interface, or an interface with DPDK support \cite{intel2014data} for high-speed packet processing. FlowFetch collects telemetry for a network flow, identified by 5-tuple: \textit{src} and \textit{dst} IP addresses, transport layer ports and protocol. In this component, multiple fully programmable telemetry functions can be associated with a flow. Two functions used in this paper are (1) \textit{\textbf{request packet counters}} and (2) \textit{\textbf{chunk telemetry}}. The first function exports the number of request packets (identified by the packet payload length) observed on the flow every 500ms. The second function (further described in \S\ref{sec:qoe}) exports metadata on isolated media chunks for estimating QoE metrics.


To isolate network flows corresponding to the video stream, FlowFetch performs a regex match on the SNI field captured in the TLS handshake of an HTTPS flow (mentioned in Table \ref{tab:vidfetch}). 
Along with network telemetry data collected for each video, the orchestrator collects playback metrics like resolution and buffer health from the video player. Twitch and YouTube expose an advanced option which displays (when enabled) an overlay with all the playback metrics. These playback metrics, stored along with the network telemetry data, will form a collocated time series dataset for each video stream.

\textbf{Campus Data Collection:} We also collected data for Twitch videos from our university campus traffic (from both WiFi and Ethernet clients). 
We received a mirror of all the traffic between campus and the Internet to one of our servers (architecture shown in Fig. ~\ref{fig:campusarch}). We employed FlowFetch to collect real user-generated Twitch live and VoD flows from various device types like personal computers, tablets, and phones. As described above, by applying SNI regex matches, FlowFetch filters and tags the collected flow as Live or VoD. Since we have no control over the device/user streaming the videos, none of the playback metrics, such as resolution and buffer health, are available. Hence, this data set can only be used for classification purposes. Table \ref{table:dataset}. shows the number of video streams collected across providers using our tool and from the campus traffic. As evident in Twitch campus data, live streams vastly outnumber VoD streams. We will also see in our field trial (\S\ref{sec:prototype}) that the majority of Twitch videos are live streaming. For our campus data collection, no limit is applied to the length of video streaming sessions. For the lab collected data, we had limited the length of browser sessions to 5 minutes.


In total, we collected over 1000 hours' worth of video playback across both providers (Twitch and YouTube). Our dataset (of video playback metrics with their corresponding network measurement) will be released when this paper is accepted. This dataset can be beneficial to the research community to understand the variations in live video stream playback in terms of resolutions, buffer sizes, latency modes, and corresponding behavioral patterns of network flows which carry the content.


\section{Classification: Live versus VoD}\label{classification}
We now design a general neural network architecture for a classifier that takes a time-series vector consisting of request packet counts. Using our collected data, we then train one instance of the classifier for each provider.

\subsection{LSTM Model Architecture}\label{sec:lstm}

We demonstrated in \S\ref{sec:characteristics} that for a network flow, the requests made for content are evidently different in live streaming compared to VoD streaming. This feature is captured in our dataset, wherein the count of requests is logged every 500ms for a given network flow. To enable real-time classification, we consider only 30 seconds of the playback as a time window over which we aim to classify the stream. We thus obtain $60$ data-points that form the input to our model as denoted by: 

\vspace{-4mm}
\begin{equation}\label{X}
\vec{X} = [x_{1},x_{2},....,x_{59},x_{60}] 
\end{equation}
\vspace{-6mm}


As we saw in Fig.~\ref{fig:timeseries}, live streams display more frequent data requests, distinguishing their network behavior across various providers. For example, in Twitch's case, data is requested every two seconds during the stable phase after initial buffering. Hence, the stable $\vec{X}$ is ideally expected to be in the form of  ``{\myverb{200020002000...}}'' -- non-zero values occurring every four data points (4 $\times$ 0.5s = 2s interval). Such patterns can be extracted by features such as \textit{zeroFrac} \ie fraction of zeros in the window, \textit{maxZeroRun} \ie maximum consecutive zeros and so on. They can then be used to train a machine-learning model. However, for different providers, the feature types and their combinations would differ significantly. Hence, instead of handcrafting features from $\vec{X}$, we aim for a classification model that derives higher-level features automatically from training data. 
Note, that unlike the lag values of top peaks in the autocorrelation function (\S\ref{sec:characteristics}) that capture limited properties of the intended signal, the $\vec{X}$ is a vector of raw time-series data, inherently capturing all temporal properties of video requests.
In order to automatically derive features of this temporal dimension, we use a popular time series model called the Long Short Term Memory (LSTM) neural network \cite{10.1162/neco.1997.9.8.1735}.

An LSTM maintains a hidden state ($\Vec{h_{t}}$) and a cell state ($\Vec{c_{t}}$), shown as upper and lower channels respectively in Fig.~\ref{fig:lstmcell}. The cell state of the LSTM acts like a memory channel, selectively remembering information that will aid in the classification task. In the context of our work, this could be the analysis of periodicity and/or the pattern by which $x_{i}$s vary over time. The hidden state of the LSTM is an output channel, selectively choosing information from the cell state required for classifying a flow as live or VoD. Fig.~\ref{fig:lstmcell} shows that at epoch $t$ the input $x_{t}$ is fed to the LSTM along with the previous hidden state $h_{t-1}$ and cell state $c_{t-1}$, obtaining current $h_{t}$ and $c_{t}$ -- at every epoch, information of the previous steps is combined with the current input. Using this mechanism, an LSTM is able to learn an entire time series sequence with all of its temporal characteristics.


As detailed above, we feed each $x_{i}$ from $\vec{X}$ into the LSTM sequentially and then obtain the final hidden state ($\vec{h_{60}}$), which retains all the necessary information for the classification task. We then feed $\vec{h_{60}}$ to a multi-layer perceptron (MLP) to make the prediction, as shown in Fig. \ref{fig:neuralnet}. The final output of the MLP is the posterior probability of the input time-series being an instance of live streaming. 

Ideally, the MLP is expected to predict a probability of $1$ when fed with an instance of live streaming and a probability of $0$ otherwise. However, in practice, we use a threshold probability of $0.5$ for predicting the flow as a live stream. In our architecture, the LSTM network has a layer consisting of a hidden vector and a cell vector, each with the size of $32\times1$, followed by an MLP with three hidden layers having dimensions of $16\times1$, $8\times1$, and $4\times1$, respectively. 
In our evaluation, we found that adding additional layers or increasing the state vector size does not improve prediction accuracy. Thus, the simplicity of our model ensures that it has a reasonably shorter training time, faster inference, and a relatively low memory footprint.

\subsection{Training and Results}
It is important to note that the neural network architecture is consistent across providers, thus highlighting the generality of our approach to classifying live and VoD streams. For the remainder of the paper, the combination of the LSTM and MLP is referred to as $model$. Although request patterns are distinct across providers, our $model$ automatically derives higher-level features from the requests data for the classification task using back-propagation and optimization techniques. While training, we create multiple mini-batches of our training data (both lab and campus data), with each batch holding $128$ streams. We then pass a batch through the $model$ and get the predicted probabilities. 
We use the binary cross-entropy loss function (BCE) 
to get the prediction error with respect to the ground truth. 
Once the error is computed, we perform backpropagation followed by Adam optimization to modify the weights in our $model$. We use a weight decay of $10^{-3}$ for the MLP weights and a learning rate $\alpha$ of $10^{-3}$.  When trained on an Nvidia GeForce GTX 1060 gpu, the size of our $model$ is 483 MB on memory.



\begin{table}[]
	\caption{\label{tab:varying} Monitoring duration impact on classification accuracy (\textit{ReCLive} $model$ vs. Random Forest).}
	\vspace{-2mm}
	\centering
	\renewcommand{\arraystretch}{1.2}	
	\begin{tabular}{l|c|c|c|c}
		\toprule
		\multicolumn{1}{l|}{}  & \multicolumn{4}{c}{Monitoring Duration}\\ \hline
		
		Provider & T=10 sec & T=20 sec  & T=30 sec & T=30 sec [\textbf{RF}]\\
		\hline
		\textbf{Twitch} & 94.33\% & 96.13\% & \textbf{96.82\%} & 89.50\% \\
		\textbf{YouTube} & 96.57\% & 98.28\% & \textbf{99.80\%} & 68.93\% \\
		\bottomrule
	\end{tabular}
\vspace{-5mm}
\end{table}

\begin{table}[]
	\caption{\label{tab:confMat} Confusion matrix of the models.}
	\vspace{-2mm}
	\centering
	\renewcommand{\arraystretch}{1.2}	
	\begin{tabular}{c|cc|cc}
		\toprule
		Provider & \multicolumn{2}{c|}{\textbf{Twitch}} & \multicolumn{2}{c}{\textbf{YouTube}}  \\ \hline
		Class & Live & VoD & Live & VoD  \\ \hline
		Live & \underline{0.981} & 0.019 & \underline{1.000} & 0.000  \\
		VoD & 0.117 & 0.883 & 0.004 & 0.996 \\
		\bottomrule
	\end{tabular}
	\vspace{-4mm}
\end{table}

With the training parameters mentioned above, the $model$ achieved high accuracies (80-20 train-test split) across both providers, as shown in Table~\ref{tab:varying}. For our baseline, we trained a Random Forest by lag values of the three highest peaks (using the auto-correlation function described in \S\ref{sec:characteristics}). 
Comparing the last column (baseline accuracies) with the other three columns (\emph{model} accuracies) in Table~\ref{tab:varying}, the performance of $model$ is far superior to that of the baseline classifier. To further understand the impact of monitoring duration on the accuracy, we quantify the performance of our $model$ with $10$, $20$, and $30$ seconds of data, as shown in Table~\ref{tab:varying}.

From these results, we make two key observations. Firstly, relatively high accuracies are seen even for the 10-sec $model$. This is mainly because the number of segments downloaded in the buffering phase of VoD and live videos differs significantly for both providers. In Twitch, VoD streams download more segments than live, whereas live streams are more active for YouTube. Such behavior is captured in the input vector $\vec{X}$ -- the start of a Twitch VoD stream looks like ``{\myverb{6101121211211200100000000...}}'' (Fig. \ref{fig:timeseriesTwitchVoD}), while a Twitch live stream displays a pattern like ``{\myverb{6202000200020002000200020...}}'' (Fig. \ref{fig:timeseriesTwitchLive}). 
Secondly, the accuracy in the $20$-sec and $30$-sec $models$ is slightly improved since live and VoD flows show evidently different patterns in their stable phase too. This is evident from the examples we saw earlier in \S\ref{sec:intro} where a Twitch live stream makes requests every 2 sec, more frequent than 10 sec in a Twitch VoD stream. 
The increased accuracy also highlights the fact that the $model$ can make better predictions when fed with larger windows of data.

Table ~\ref{tab:confMat} shows the confusion matrix of our $30$-sec $model$ across providers. We observe almost perfect true positive rates (underlined values) across Twitch and YouTube for live flows. However, for VoD flows, we observe a lower performance in Twitch. We believe this is because the Twitch data consists of real-user-generated streams (collected from the production network of our university campus). In contrast, the YouTube data was selectively collected in a lab environment. In particular, we found certain instances of Twitch VoD in low-bandwidth conditions where the client makes numerous video requests, resulting in an input $\vec{X}$ that is similar to a window of a live stream.

\section{Estimating QoE metrics of Live Video} \label{sec:qoe}
While the QoE of a live video stream is subjective, we capture it with two major \textit{objective} metrics; video quality and buffer stalls. Video quality can be measured using: (a) resolution of the video, (b) bitrate (number of bits transferred per sec), and (c) more complex perceptual metrics like MOS \cite{winkler2008evolution} and VMAF \cite{li2018vmaf}. In this paper, we develop a method to estimate the resolution of the playback video since the ground-truth data is available across both providers. Also, the resolution is typically reported (or available to select) in popular live streaming services. In addition to video resolution, we devise a method to detect the presence of buffer stalls that are more likely to occur in live streaming (compared to VoD) since a smaller buffer size is maintained on the client to reduce the latency between the producer and the viewer. In what follows, we present our analysis of data collected from the network consisting of audio/video segments versus metrics recorded on the client. Subsequently, we develop methods that estimate video resolution and detect buffer stalls.

\subsection{Network-Level Measurement}
To estimate QoE metrics for the live stream, we need to estimate the size of the media segments being fetched. The amount of data downloaded between two consecutive requests reasonably estimates the size of media segments. We refer to this estimate as a \emph{chunk}. Hence, we use the term \textit{segment} for a unit of media requested by the player. In contrast, \textit{chunk} denotes a corresponding unit of data observed on the network (demarcated by the request packets).
We build upon existing network chunk-detection algorithms \cite{gutterman2019requet,mangla2018emimic} to isolate the video chunks fetched by the live player. In short, the algorithm identifies the start of a chunk by an upstream request packet and aggregates all subsequent downstream packets to ``form'' the chunk. For each chunk, it extracts the following features: \textit{\textbf{requestTime}}, \ie the timestamp of the request packet, \textit{\textbf{requestPacketLength}}, \textit{\textbf{chunkStartTime}} and \textit{\textbf{chunkEndTime}}, \ie timestamps of the first and the last downstream packets following the request, and lastly \textit{\textbf{chunkPackets}} and \textit{\textbf{chunkBytes}}, \ie total count and volume of downstream packets corresponding to the chunk. 

During the playback of a live video stream, the chunk telemetry function operates on a per-flow basis in our FlowFetch tool. It exports the above features for every chunk observed on the five-tuple flow(s) carrying the video. In addition, as earlier mentioned in \S\ref{sec:dataset} we collect resolution and buffer health metrics reported by the video client. In what follows, we correlate and analyze the chunk data obtained from the network and client metrics to train our models for estimating resolution and detecting the presence of buffer stalls.

\subsection{Estimating Resolution}
The resolution of a live video stream indicates the frame size of video playback -- it may also sometimes indicate the rate of frames being played. For example, a resolution of 720p60 means the frame size is 1280$\times$720 pixels while playing 60 frames per sec. For a given fixed duration video segment, the video segment size  (and hence our corresponding chunk estimate) usually increases in higher resolutions as more bits need to be packed into the segment. 


We analyzed the live video streams played using our tool for both content providers to better understand the distribution of video segment sizes across various resolutions. We also consider four bins of resolution, namely Low Definition (LD), Standard Definition (SD), High Definition (HD), and Source (originally uploaded video with no compression, only available in Twitch) -- Table~\ref{tab:resdist} shows the distribution of streams across these bins. The bins are mapped as follows, anything less than 360p is LD, 360p and 480p belong to SD, 720p and beyond belongs to HD. If the client tags a resolution (usually 720p or 1080p) as Source, it is binned into Source. Such binning serves two purposes: (a) it accounts for a similar visual experience for a user in neighboring resolutions, and (b) it provides a consistent way to analyze across providers. Fig.~\ref{fig:reschunks} shows the distribution of chunk sizes versus resolutions, and will be further explained later in \S\ref{sec:analinfer}. We estimate the resolution in two steps: (a) first, separating chunks corresponding to video segments, and (b) next developing an ML-based model to map the chunk size to resolution.

\begin{table*}
	\begin{minipage}{0.7\linewidth}
		\centering
		\includegraphics[width=0.45\textwidth,height=0.33\textwidth]{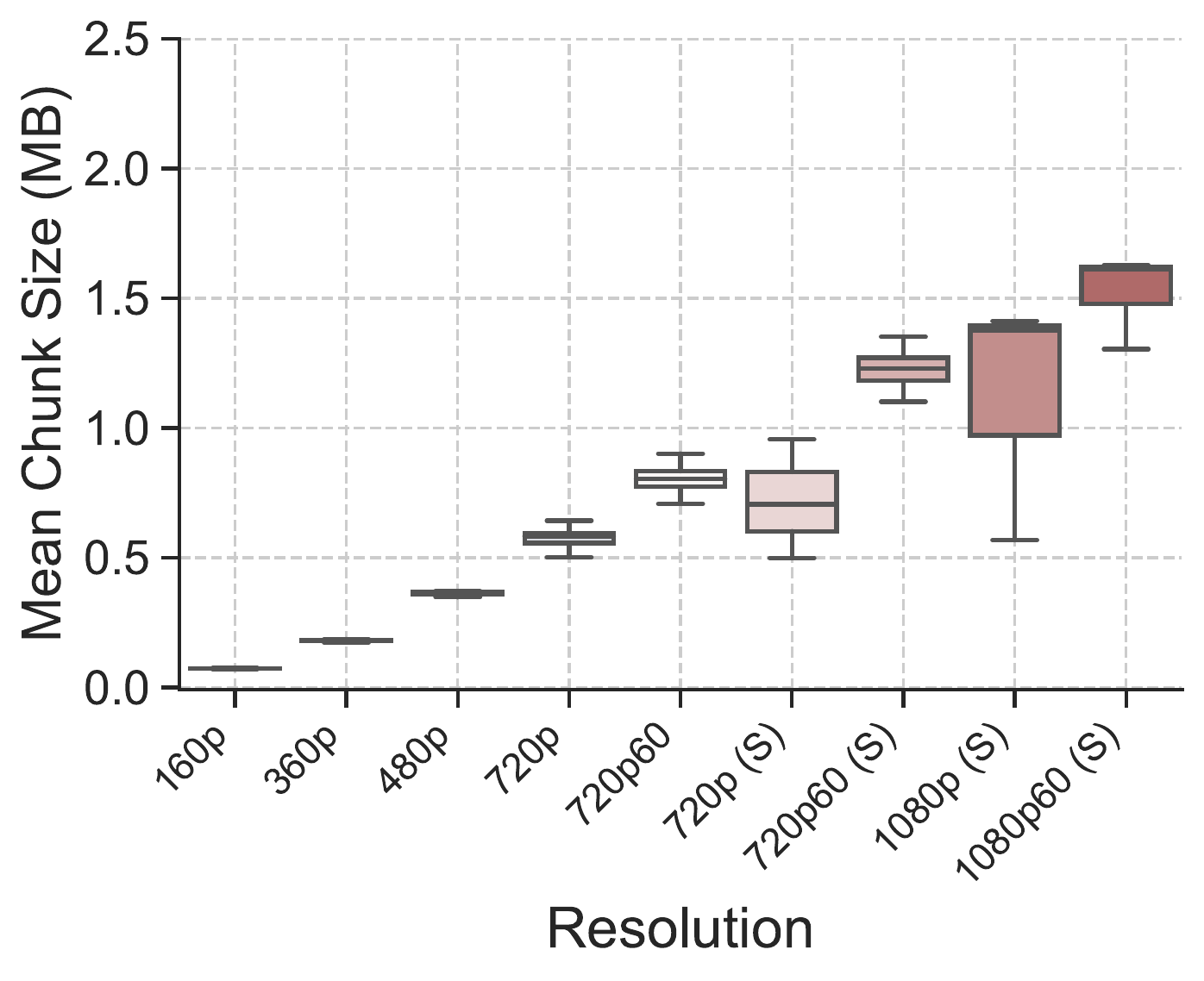}
		\label{fig:resExactTW}
		\hspace{5mm}
		\includegraphics[width=0.45\textwidth,height=0.33\textwidth]{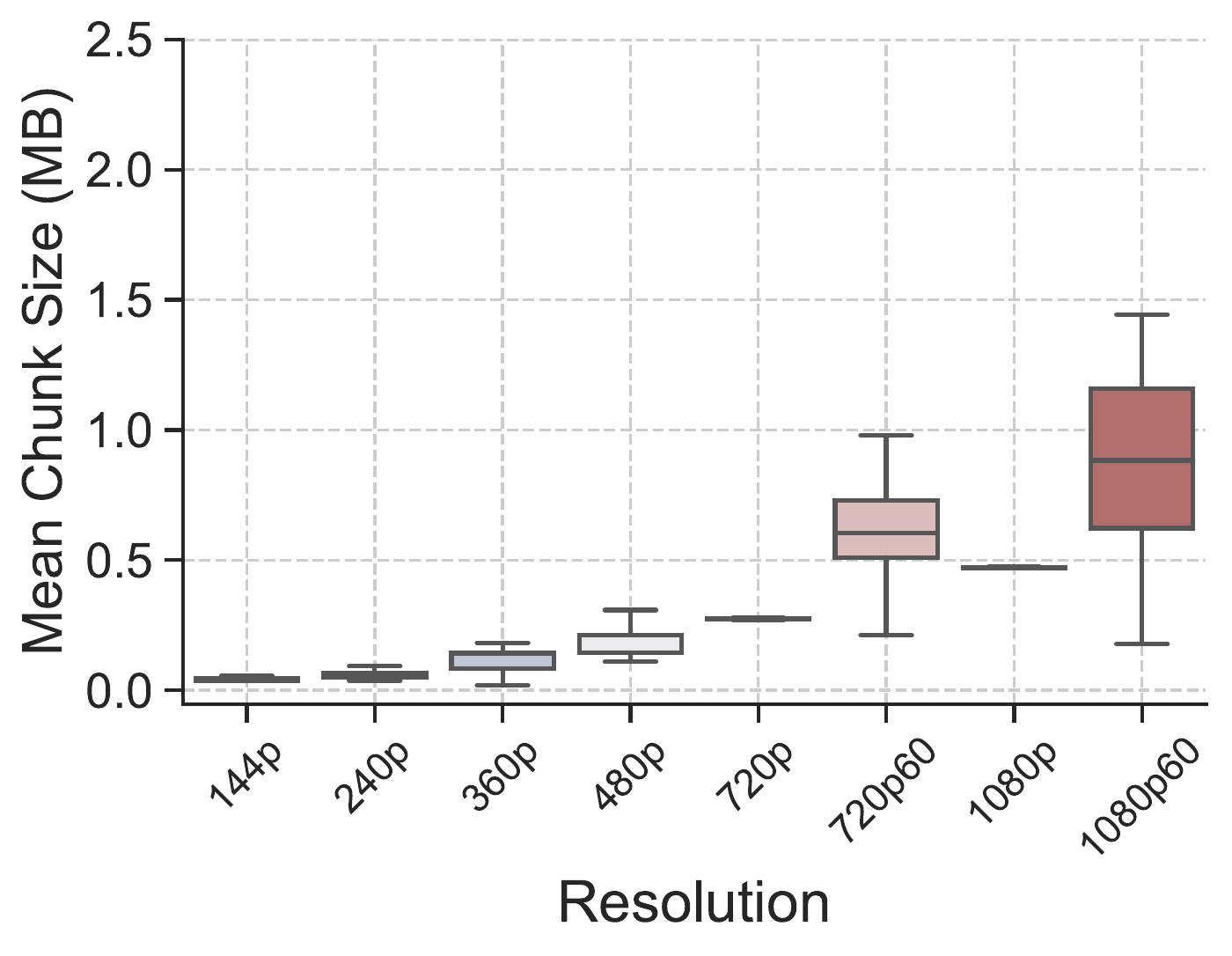}
		\label{fig:resExactYT}
		\par
		\vspace{\baselineskip}
		\includegraphics[width=0.45\textwidth,height=0.33\textwidth]{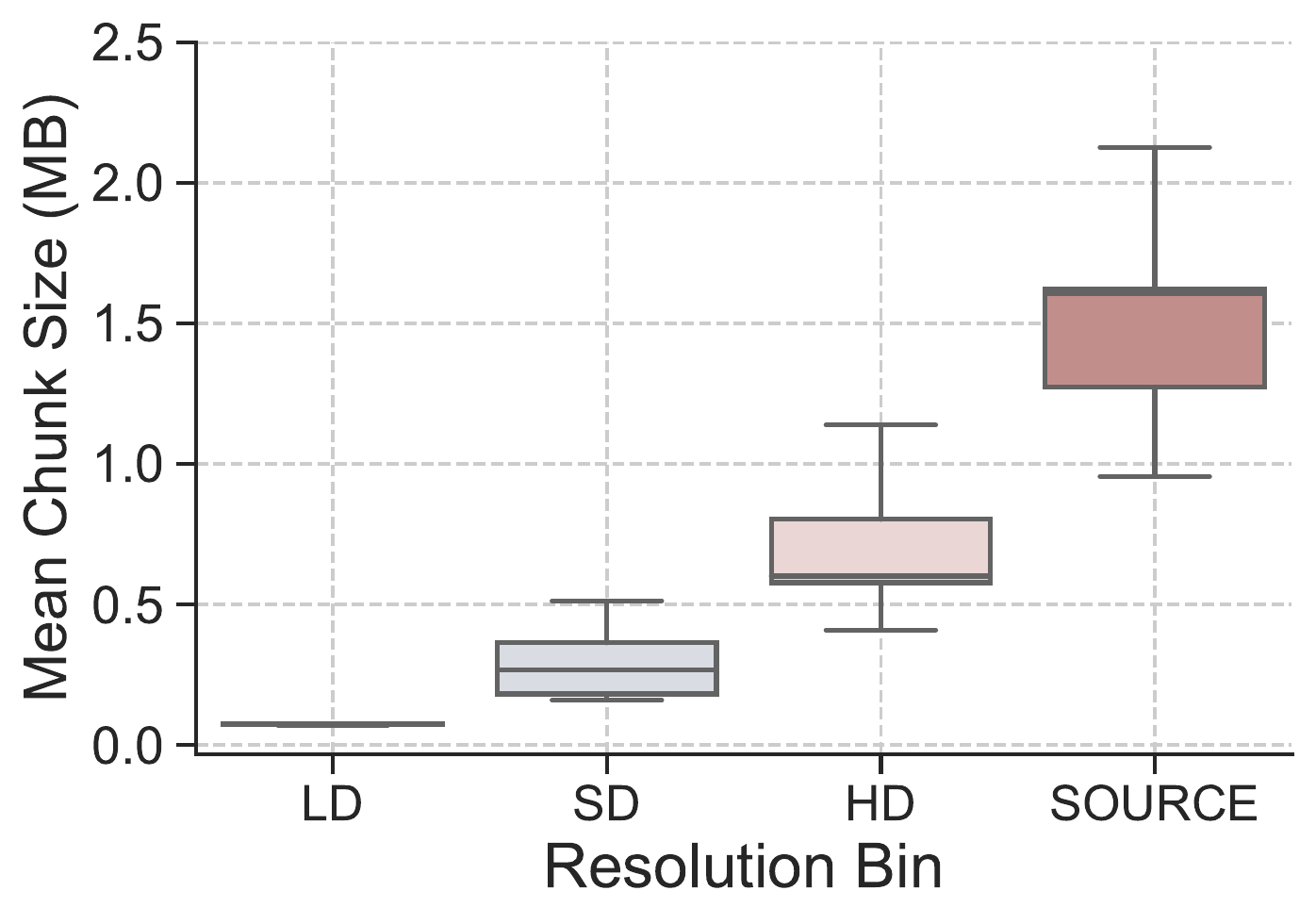}
		\label{fig:resLevTW}
		\hspace{5mm}
		\includegraphics[width=0.45\textwidth,height=0.33\textwidth]{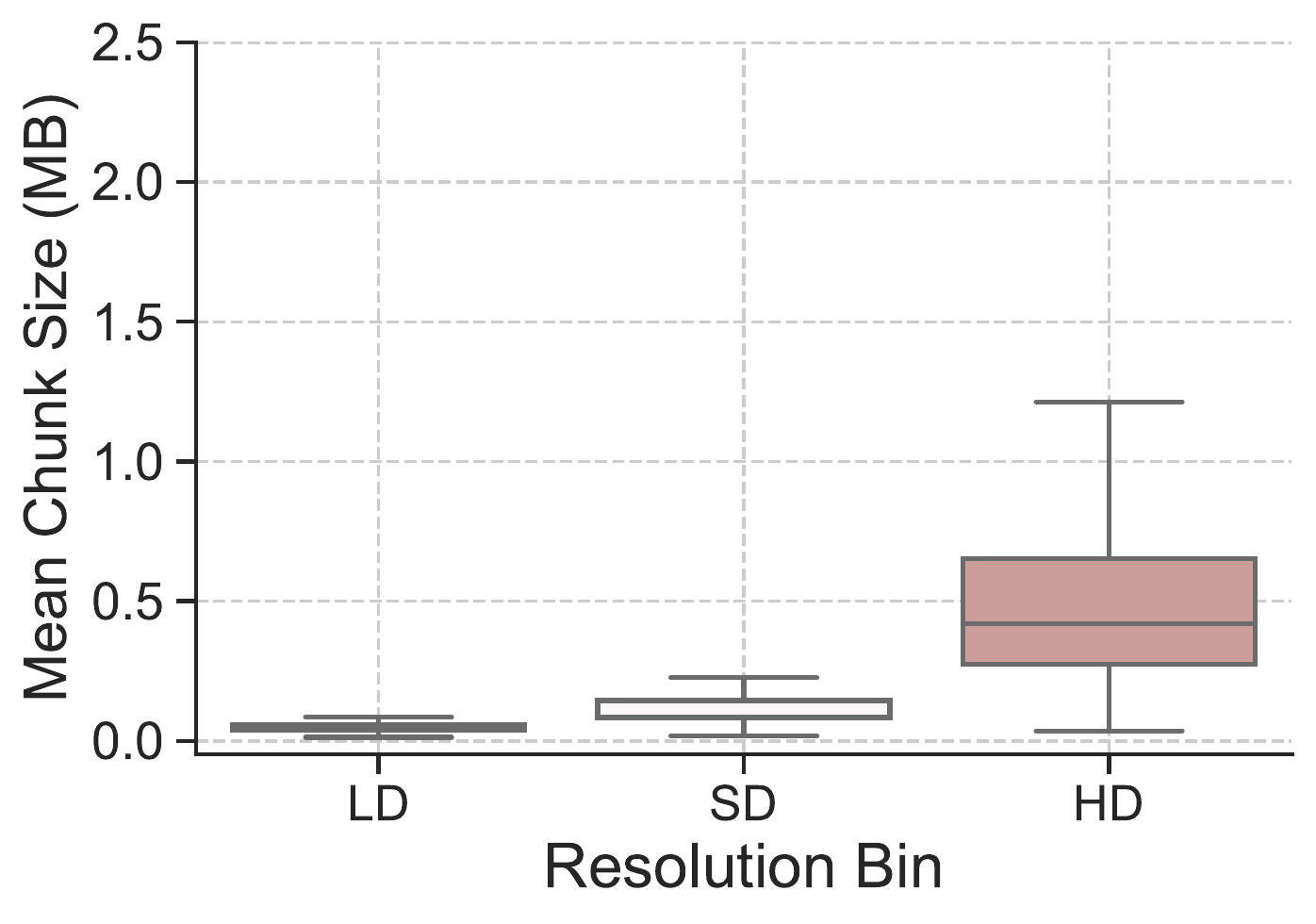}
		\label{fig:resLevYT}
		\captionof{figure}{Chunk size versus resolution for Twitch (left), and YouTube (right).}
		\label{fig:reschunks}
		\vspace{-4mm}
	\end{minipage}\hfill
	\begin{minipage}{0.33\linewidth}
		\centering
		\caption{\label{tab:resdist} Dataset resolution distribution.}
		\begin{adjustbox}{max width=0.99\textwidth}	
			
			\begin{tabular}{@{}lcccc@{}}
				\toprule
				Provider & LD   & SD   & HD   & SOURCE \\ \midrule
				Twitch   & 17\% & 32\% & 34\% & 17\%   \\
				YouTube  & 36\% & 36\% & 28\% & -      \\ \bottomrule
			\end{tabular}
		\end{adjustbox}
		\vspace{\baselineskip}
		\vspace{25mm}
		\caption{\label{tab:resacc} Resolution prediction accuracy.}
		\begin{adjustbox}{max width=0.99\textwidth}
			\begin{tabular}{@{}lcc@{}}
				\toprule
				Provider & Resolution & Resolution bin \\ \midrule
				Twitch & 90.64\% & \textbf{97.62}\% \\ 
				YouTube & 75.17\% & \textbf{90.08}\% \\ \bottomrule
			\end{tabular}
		\end{adjustbox}
	\end{minipage}
\end{table*}

\subsubsection{Separation of video chunks}\label{subsec:vidsep}
Network flows corresponding to a live stream can carry chunks of data that correspond to any of video segments, audio segments, or manifest files, and hence the video component needs to be separated out to estimate its resolution. 
However, the method to isolate video segments depends on the provider -- it can be developed by analyzing a few examples of streaming sessions and/or by decrypting SSL connections and analyzing the request URLs.

\textbf{Twitch} usually streams both audio and video segments on the same 5-tuple flow for live video streaming, and manifest files are fetched in a separate flow. We observed that audio is encoded in a fixed bitrate, and thus its chunk size is consistent ($\approx 35$ KB). Further, Twitch video chunks of the lowest available bitrate (160p) have a mean of $76$ KB. Thus, video chunk identification is fairly simple for Twitch live streams, \ie all chunks more than $40$ KB in size.


\textbf{YouTube} live usually uses multiple TCP/QUIC flows to stream the content consisting of audio and video segments  --  Youtube operates manifest-less. As indicated in Table \ref{tab:vidfetch}, Youtube live operates in two modes, \ie Low Latency (LL) with 2 sec periodicity of content fetch, and Ultra Low Latency (ULL) with 1 sec periodicity. We found that the audio segments have a fixed bitrate (i.e., size per second is relatively constant) regardless of the latency mode -- audio chunk size of $28-34$ KB for the ULL mode, and $56-68$ KB for the LL mode. However, separating the video chunks is still nontrivial as video chunks of 144p and 240p sometimes tend to be smaller in size than the audio chunks.

To separate the audio chunks, authors of \cite{gutterman2019requet} used the \textit{requestPacketLength} as they observed that the audio segment requests were always smaller than the video requests. We used this method for TCP flows but found it inaccurate in UDP QUIC flows as the audio segment requests are sometimes larger than video segment requests due to header compression.
Further, QUIC flows pose additional challenges, especially for live video streams. Because of bi-directional stream support available in HTTP/2 + QUIC, a request for a media segment can be sent before the previous segment completely downloads. Since our chunk telemetry function relies on the request packets to mark the start of chunks, the \textit{chunk} sizes computed by the network telemetry function differ from the actual size of media \textit{segments}. For this reason, we cannot accurately capture individual media segments for YouTube videos delivered over QUIC flows. Thus, while we detect QUIC live video (as the request patterns are still distinguishable), QoE inferencing for YouTube QUIC video streams is beyond the scope of this paper. 

\subsubsection{Analysis and Inference}\label{sec:analinfer}

After identifying the chunks corresponding to the video segments for each provider, we now look at the distribution of chunk sizes across various resolutions at which the video is played. Fig.~\ref{fig:reschunks} shows box plots of mean (video) chunk size in MB versus the resolution (\ie actual value or binned value) in categorical values. Note that the mean chunk size is computed for individual video streams of duration 2-5 minutes. Further, the label \textit{(S)} on the X-axis indicates a Source resolution. 

Looking at Fig.~\ref{fig:reschunks}, we make the following observations: (a) video chunk size increases with resolution across both the providers; (b) chunk sizes are less spread in lower resolutions; and (c) chunk sizes of various transcoded resolutions (\ie not the source resolution) do not overlap much with each other for Twitch. However, the overlap of neighboring resolutions becomes more evident in YouTube streams. Such overlaps make it challenging to estimate the resolution.

We use the Random Forest algorithm for mapping chunk sizes to the resolution of playback 
as it creates overlapping decision boundaries using multiple trees and then uses majority voting to estimate the best possible resolution by learning the distribution from the training data.
Using the mean chunk size as an input feature, we trained two models, \ie one estimating the exact resolution and the other estimating the resolution bin. We perform 5-fold cross-validation on the dataset with 80-20 train-test split, and our results are shown in Table \ref{tab:resacc}.

\subsection{Predicting Buffer Stalls}
Buffer stalls occur when the playback buffer is emptied out because the video segments cannot be fetched in time. This QoE metric is vital for live streams, which typically maintain short buffers (4 seconds for Twitch LL and Youtube ULL modes). 
Even for a few seconds, network instability can cause the live buffer to deplete, leading to a stall causing viewer frustration.

To better understand the live buffering mechanism across the three providers, we collect data for live video streams ($\approx 5 min$ per session) while using the network conditioner component of our tool to impose synthetic bandwidth caps. We created a commonly occurring situation in a household wherein cross-traffic (browsing/e-mail etc.) is introduced for a few seconds while a live stream is going on. To do so, the tool starts with a cap of 10 Mbps (typical household bandwidth) and then, after every 30 seconds, caps the download/upload bandwidth at a random value (between 100 Kbps to 2 Mbps) for a duration of 10 seconds (mimicking the congestion due to cross traffic). Live videos being played in the browser are accordingly affected by these bandwidth switches. We found that if videos are played at \textit{Auto} resolution, then the clients avoid stalls most of the time by switching to lower resolutions. Therefore, we forced the video streams to play at one of the HD resolutions (1080p or 720p) to gather data of buffer stall events. 
In total, we collected more than 250 video streams across the three providers. On average, 15\%  and 6\% of the playback time were spent in the stall state for Twitch and YouTube.

\begin{figure}[t!]
	
	\begin{center}
		\includegraphics[width=0.95\columnwidth]{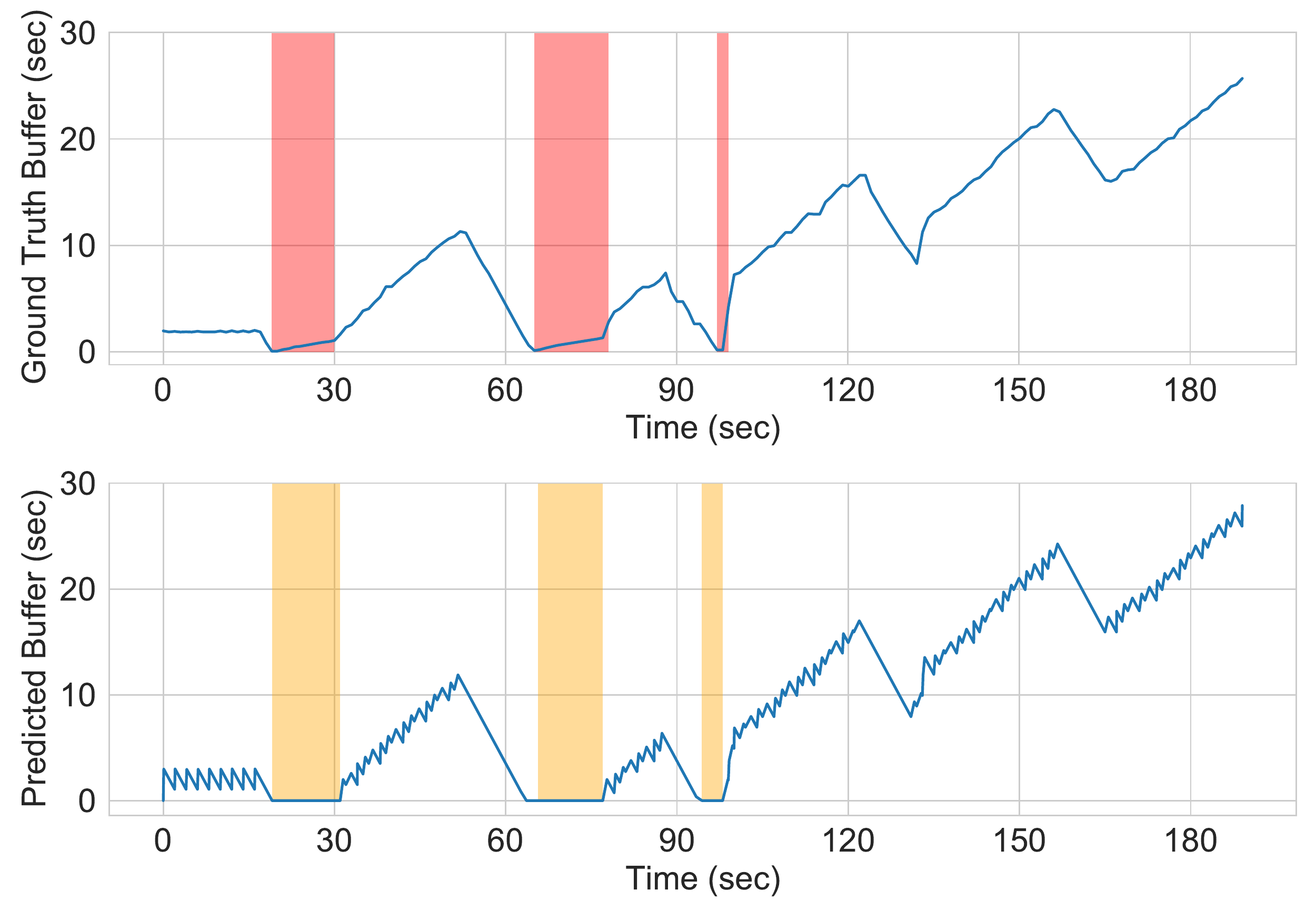}
		\vspace{-2mm}
		\caption{\label{fig:bufferpredict} Time-trace of buffer health value: ground-truth  predicted, for a sample Twitch stream.}
		\vspace{-6mm}
	\end{center}
\end{figure}

Fig. \ref{fig:bufferpredict} shows the dynamics of buffer health for a representative stream in our dataset. We observe that this low latency Twitch video starts with 2 seconds of the buffer. It soon encounters the first stall (highlighted by the red bar) around second 25 due to network congestion caused by cross traffic. Following that, the stream linearly increases its buffer to 10 seconds but experiences stalls a couple more times until the second 100. After this point, the buffer value increases to more than 20 seconds. It can be seen that a stall event not only deteriorates user experience but also increases the latency of the live stream as the user is watching content that was recorded at least 20 seconds ago -- defeating the purpose of live streaming. 

To predict such stalls, our buffer estimator algorithm takes two parameters as input. The first is $\mbox{\textit{Seg}}_{dur}$; live video streams typically encode content into video segments of fixed duration. This duration depends on the playback mode - for instance, YouTube ULL streams have $\mbox{\textit{Seg}}_{dur}=1$ sec, while YouTube LL streams have $\mbox{\textit{Seg}}_{dur}=2$ sec (\S\ref{sec:characteristics}). We automated this estimation by equating $\mbox{\textit{Seg}}_{dur}$ to be the median inter-request time (IRT) of video segments in the first window of $n$ seconds (empirically configured to be 20 sec). The second parameter is $\mbox{\textit{Buf}}_{min}$;  a client typically fetches few video segments (at least one) until a minimum buffer is filled before it begins playback. In the case of Twitch, playback begins after the first segment finishes downloading -- hence, $\mbox{\textit{Buf}}_{min}=2$ sec (one segment long). However, in the case of YouTube, $\mbox{\textit{Buf}}_{min}$ seemed to vary between 2-10 seconds. Thus, we conservatively choose the mean value in our dataset --  3 sec for ULL streams and 6 sec for LL streams. 

Using the parameters above and the isolated video chunks mentioned above (\S\ref{subsec:vidsep}), the buffer estimation algorithm (Algorithm \ref{buffalgo}) works as follows. At the beginning of a stream, its buffer is initialized at zero and increases by steps of $\mbox{\textit{Seg}}_{dur}$ at the end of every chunk observed on the network, until it reaches $\mbox{\textit{Buf}}_{min}$ (Algorithm \ref{buffalgo}, Lines: 4-7). For every subsequent video chunk, the buffer value is adjusted by: (a) adding $\mbox{\textit{Seg}}_{dur}$ and (b) subtracting the time elapsed in the playback since its previous chunk (Algorithm \ref{buffalgo}, Lines: 4,8).

\begin{algorithm}[t]
	\caption{Predict Buffer Stall.}
	\label{buffalgo}
	\DontPrintSemicolon 
	\SetNoFillComment
	\Parameter{$\mbox{\textit{Buf}}_{min}, \mbox{\textit{Seg}}_{dur}$}
	\KwData{\textit{Chunks} detected on network $\{c_1,c_2, \ldots, c_n\}$}
	\KwOut{Estimated buffer health}
	$b \gets 0$ \Comment*[r]{Tracks current buffer value}
	$t \gets 0$ \Comment*[r]{Tracks endTime of last chunk}
	\For{each network chunk c} {
		b  += $\mbox{\textit{Seg}}_{dur}$ \\  
		\If{$b <= \mbox{\textit{Buf}}_{min}$}  {
			$t  = c.EndTime$\;
			\Continue  \Comment*[r]{Wait until video starts}
		}
		b --= $c.EndTime - t$ \Comment*[r]{Decrement buffer by time elapsed}
		\If{$b <= 0$}{
			$b = 0$\Comment*[r]{Stall detected}
		}
		$t  = c.EndTime$\;
	}
	\vspace{-1mm}
\end{algorithm}

Our algorithm predicts the current buffer value (in seconds) of the client video player for both providers. To quantify the accuracy of predicting buffer stalls (buffer value = 0), we first divide a given video stream into 5-sec windows and assign a boolean value (true when there was a stall and false otherwise) to each window. The ground truth of buffer stalls comes from the playback metrics collected by our tool described in \S\ref{sec:dataset}. Table~\ref{tab:stallresults} summarizes the performance of predicting buffer stalls across all playback windows. Among all the windows across the video playback, we computed the accuracy, precision and false positive rate of our algorithm across providers as shown in Table~\ref{tab:stallresults}.
Overall, our algorithm yields about $90$\% accuracy in predicting the presence of a buffer stall in a 5-sec window. 
Note that our method tends to underestimate the buffer health in YouTube videos (false positive rate $14.2$\%)  since we choose a conservative $\mbox{\textit{Buf}}_{min}$ value, leading us to predict stalls even when the buffer value is small but non-zero.
We found that in more than 50\% of the false-positives, our algorithm underestimates the buffer value by at most $2.3$ sec ($\approx$ the duration of a segment). 


\section{Prototype Implementation and \\Field Evaluation} \label{sec:prototype}

We prototyped our scheme and deployed it in an ISP network serving over 7000 home subscribers. In what follows, we describe the architecture of our system, insights gathered from the field trial which led to the development of a state machine to make our classification component more robust. Further, we discuss various practical challenges encountered during the deployment and highlight the possible future directions of the work.

\begin{table}[t]
	\caption{\label{tab:stallresults} Buffer stall prediction results.}
	\centering
	\begin{tabular}{@{}llll@{}}
		\toprule
		& Accuracy & Recall & FP rate \\ \midrule
		Twitch  & 90.1\%   & 90.0\%      & 10.3\%    \\
		YouTube & 89.6\%     & 88.2\%      & 14.2\%    \\ \bottomrule
	\end{tabular}
	\vspace{-3mm}
\end{table}

\begin{figure}[t]
	\centering
	\includegraphics[width=0.485\textwidth]{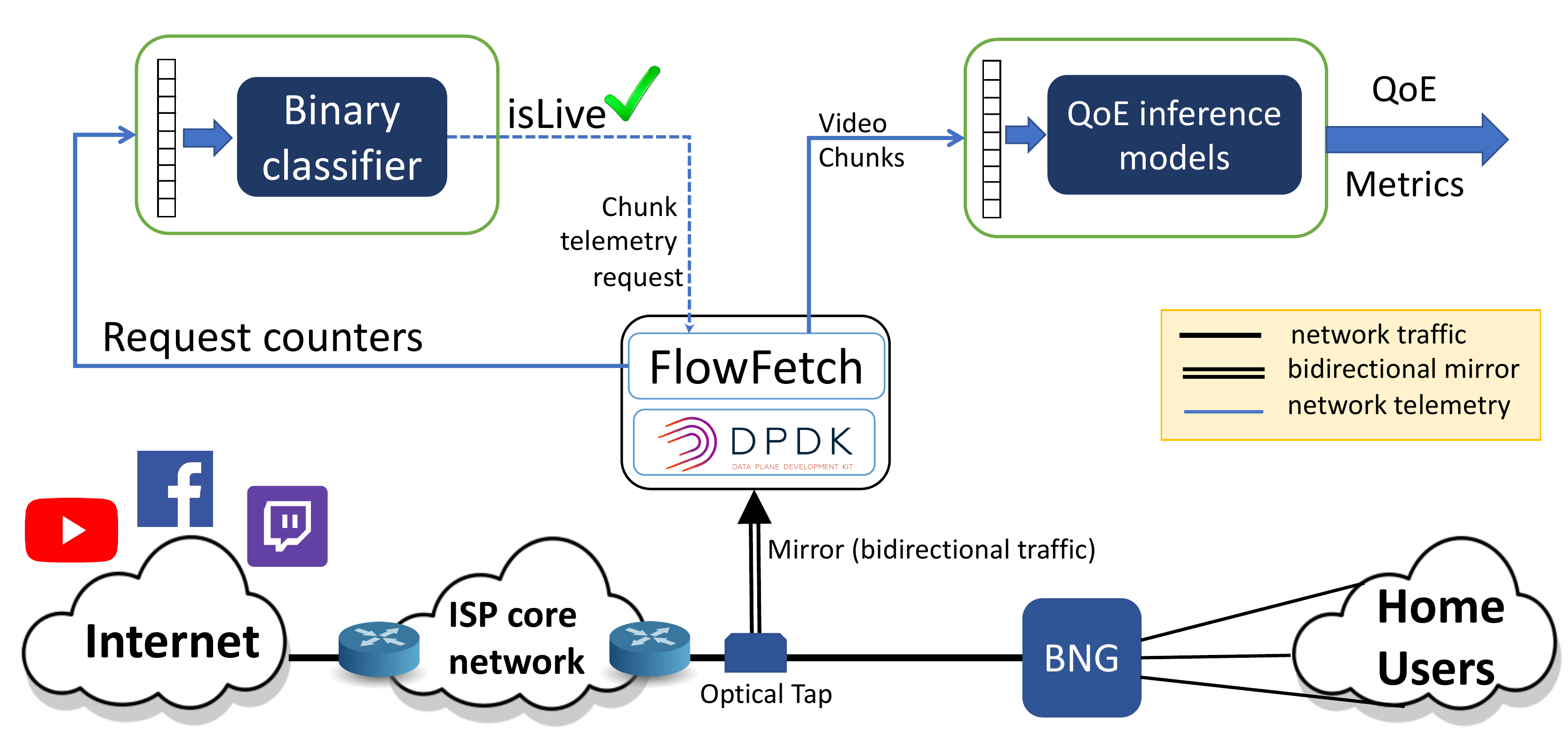}
	\vspace{-3mm}
	\caption{Architecture of our prototype.}
	\label{fig:deparch}
\end{figure}

\subsection{Prototype Architecture}\label{sec:deploy_arch}

\begin{table*}[t]
	\caption{Summary of our field evaluation data.}
	\vspace{-2mm}	
	\centering
	\label{tab:usereng}
	\renewcommand{\arraystretch}{1.2}
	\begin{tabular}{l|cr|cr|cr}
		\toprule
		Provider & \multicolumn{2}{c|}{\# subscribers} & \multicolumn{2}{c|}{\# hours [\textit{daily avg.}]}      & \multicolumn{2}{c}{\# streams {[}\textit{avg. duration of stream}{]}}        \\ \hline
		& Live & \multicolumn{1}{c|}{VoD} & Live & \multicolumn{1}{c|}{VoD} & Live & \multicolumn{1}{c}{VoD} \\
		Twitch   & \multicolumn{1}{r}{757}    & 47     & \multicolumn{1}{r}{5,422 [\textit{774}]} & 10 [\textit{1.4}]     & \multicolumn{1}{r}{27,810 {[}\textit{11.7 min}{]}} & 54  {[}\textit{11.4 min}{]}     \\
		YouTube  & \multicolumn{1}{r}{1,438}  & 5,362  & \multicolumn{1}{r}{1,440 [\textit{205}]} & 61,432 [\textit{8,776}] & \multicolumn{1}{r}{5,801 [\textit{14.9 min}]}  & 376,120 [\textit{9.8 min}] \\ \bottomrule
	\end{tabular}
\vspace{-5mm}
\end{table*}

The architecture of our prototype deployment is shown in Fig. \ref{fig:deparch}. The ISP installed an optical tap between their core network and a Broadband Network Gateway (BNG), that aggregates traffic from residences in a particular neighborhood. 

Our system works off this tap traffic, thereby receiving a copy of every packet to/from these residences without introducing any risk to the operational network. Upstream and downstream traffic is received on separate optical tap links, and the aggregate bidirectional rate was observed to be no more than 8 Gbps even during peak hours. The traffic is processed by a Linux server running Ubuntu 18.04 with DPDK support for high-speed packet processing. Our tool FlowFetch interacts with DPDK to fetch raw packets and executes telemetry functions to export request packet counters and chunk features. Since we used the same tool during the training phase, we test our models on this data without additional pre-processing. 

	The flow of events in our system is as follows: First, we detect flows carrying video streams originating from Twitch and YouTube by performing pattern matches on the SNI field present in the TLS handshake (\S\ref{sec:netanalysis}). Every such flow is allocated the first telemetry function (\S\ref{sec:dataset}), which exports the request packet counts every 500 ms. This data is batched up in time (30-sec window -- \S\ref{sec:lstm}) to form the input vector for our LSTM-classifiers. Using the SNI match, the provider-specific model is called on this request vector. The classification is reported back to the FlowFetch component, which subsequently updates the telemetry function -- if a live stream, the second telemetry function (\S\ref{sec:dataset})  is attached to the same flow to start exporting chunk features (\S\ref{sec:qoe}) to compute the QoE metrics; if a VoD stream, telemetry functions are turned off. 
	
	\textbf{Deploying real-time classification}: Firstly, note that for Twitch, SNIs are used as ground-truth data to distinguish live from VoD streams, and hence we are able to quantify the accuracy of the Twitch classifier model. For YouTube, however, in the absence of distinct SNIs (ground-truth data), we simply rely on the model's prediction to label a window as live or VoD, and hence the model performance cannot be evaluated in the field trial. Secondly, for YouTube, the two-tuple identifier consisting of \textit{ClientIP and Protocol} was used to aggregate all the 5-tuple flows that constitute a video stream. For example, given a clientIP watching a YouTube video stream, the player might use 3-4 flows to fetch content associated with this stream. 
	Flow aggregation is not required for Twitch as the entire content of each video stream is delivered on a single flow. 
	Thirdly, note that Twitch uses an independent flow for each video stream (switching to a new video stream would generate a new flow). Hence, we classify Twitch streams ``only for their first window'' and use the prediction to tag the entire stream accordingly. YouTube streams, however, tend to re-use flows across consecutive video sessions (whenever the content is sourced from the same CDN server), even when the user transitions from a VoD stream to a live stream -- flows that carry VoD traffic may start carrying live video traffic. For this reason, the model needs to ``continuously'' classify YouTube flows during the entire lifetime of the stream (explained in \S\ref{sec:class_dep}).
	
	\textbf{Deploying real-time QoE prediction}: In order to report real-time QoE metrics, we batch up the video chunks for a window of suitable size (discussed later). We then proceed to estimate resolution and predict  whether buffer stalls occur in that window, or not. As explained in \S\ref{sec:qoe}, we first isolate the video chunks with a provider-specific method, next compute the mean chunk size of the window, and then pass it on to its provider-specific random forest classifier, which predicts the resolution bin. For our field trial, we chose to predict the resolution bin (rather than the exact resolution). It gives better accuracy and presents a consistent view of QoE across providers. The same window of chunks is passed on to the algorithm that detects buffer stalls. The predicted resolution bins and buffer stalls are then stored in a database which can be visualized in real-time or post-processed for network resource provisioning. 
	
	We note that the window length is a parameter that needs to be chosen by the network operator considering the trade-off between system response time and the accuracy of predicted resolution bins. 
	A large window size (say 60 seconds) evens out the variability in the chunk sizes, making the predicted resolution bins more accurate at the expense of making the system relatively less responsive. Similarly, a small window size (say 10 seconds) makes the system more responsive at the expense of decreased accuracy. 
	We empirically tuned our system window length to 30 seconds since it allowed for an acceptable system response time and ensured that enough chunks were captured to make a reasonably accurate prediction. A similar window length was also used in prior work \cite{bronzino2019inferring}.


\begin{table*}[t!]
	\caption{Distribution of states for YouTube streams.}
	\centering
	\label{tab:you_state}
	\renewcommand{\arraystretch}{1.2}	
	\begin{tabular}{@{}l|r|rrrr@{}}
		\toprule
		Subset       & {Distribution} & \multicolumn{4}{c}{Time fraction spent in each  \textit{State}}                \\ 
		&          &\emph{Surely VoD} & \emph{Maybe VoD} & \emph{Maybe Live} & \emph{Surely Live} \\ \hline
		All sessions     & 381,921 [\textit{100\%} ]  & 97.4\%     & 0.8\%     & 0.4\%      & 1.4\%       \\ \hline
		Only VoD         & 348,190 [\textit{91.2\%}]  & 100.0\%       & -        & -         & -          \\
		Only Live        & 1,430 [\textit{  0.4\%}]  & -        & -       & -        & 100.0\%        \\
		Stray Live       & 24,588 [\textit{  6.4\%}]  & 96.6\%        & 2.4\%      & 0.5\%       & 0.5\%          \\
		Consecutive Live & 7,713 [\textit{  2.0\%}]  & 75.7\%        & 5.7\%      & 3.7\%       & 14.9\%       \\ \bottomrule
	\end{tabular}
	\vspace{2mm}
\end{table*}

\subsection{Classifier Deployment} \label{sec:class_dep}
We successfully deployed our classification models in the partner ISP network for a trial period of one week (from 11th to 17th of July 2020). In total, our models analyzed network activity of more than 370,000 video streams, spanning over 65,000 hours of video playback viewed by more than 8,500 unique IP addresses (IPv4 and IPv6 of over 7000 subscribers). 


Table \ref{tab:usereng} summarizes our findings from the field evaluation data. We observe how subscribers of this network consume live and VoD streams from Twitch and YouTube. It can be seen in the second column that $94$\% of Twitch users watch live streams while this measure is about $21$\% for YouTube users -- this is not surprising since the Twitch platform is predominantly used for live game video streaming, and YouTube is the most popular VoD platform in many countries worldwide while its live streaming service has recently started gaining traction.
This contrast in service popularity is also substantiated at stream level (rightmost column), $99.8\%$ of Twitch streams are classified as live, and $98$\% of YouTube sessions are VoD.
Moreover, from our field evaluation data, we make the following observations: (a) around 10\% of the network subscribers consume live content on Twitch (these subscribers are probably online gamers), and approximately 20\% of the subscribers watch YouTube Live, (b) live and VoD content on Twitch are watched for about the same duration ($\approx$11 minutes)  -- this is because VoD content on Twitch consists of recorded live gaming sessions, and hence has similar viewing patterns, (c) user engagement with YouTube Live seems better (50\% longer on average) in comparison with YouTube VoD. These observations highlight a growing trend in live content engagement (probably due to growing live sporting events that run for hours). 

As explained in \S\ref{sec:characteristics}, SNI can be used as ground-truth data for Twitch services. Therefore, we evaluated the efficacy of our model in predicting Twitch video streams and found our accuracy to be $94.57$\%. Misclassifications were predominantly live streams classified as VoD -- when manually inspecting their network activity, we found that their pattern surprisingly matched that of a VoD stream with idle periods. 
As mentioned in \S\ref{sec:deploy_arch}, Twitch content is streamed over one TCP flow, and hence a one-off classification suffices to detect live streams. However, for YouTube,  we employed a classification architecture that continuously classifies a group of flows originating from a clientIP.  While many sessions were continuously classified as \textit{VoD} or \textit{Live}, there were instances where one viewing session switched from \textit{VoD} to \textit{Live}. In what follows, we draw more insights into these sessions by further analyzing the prediction results for YouTube videos. Note that for YouTube, we use the term \textit{session} to refer to a continuous watching period since identifying individual video streams is non-trivial. 

\subsection{YouTube State Machine}\label{sec:state_machine}

Based on the model prediction for individual windows (of network activity) of a session, we categorize YouTube sessions into four subsets: (a) \emph{Only VoD}, (b) \emph{Only Live}, (c) \emph{Stray Live}, and (d) \emph{Consecutive Live}, as shown by rows of Table~\ref{tab:you_state}. The subsets \emph{Only VoD} and \emph{Only Live} represent those sessions that have been consistently (during their entire lifetime) classified as VoD and live, respectively. Unsurprisingly, a majority ($91.2\%$) of the sessions are predicted as \emph{Only VoD}, while \emph{Only Live} sessions are at an extreme minority ($0.4\%$). The remaining sessions are \emph{mixed} \ie parts of their playback have been predicted as live, and other parts as VoD. We further bifurcate the \emph{mixed} YouTube sessions into two subsets: \emph{Stray Live} and \emph{Consecutive Live}. \emph{Stray Live} sessions contain a live window (stray) surrounded by VoD windows on both sides. For example, a session with a predicted pattern of windows like ``{\myverb{[VoD,VoD,VoD,live,VoD,VoD]}}'' is considered to be stray live. Note that the likelihood of a user switching between services so rapidly is fairly low, and hence we deem this live prediction as a false positive. 
We attribute these false positives to trick-play actions (fast-forward and/or rewind) triggered by a VoD-watching user that forces the player to download a large number of chunks, thus making it similar to a stable phase window of a YouTube live session (we address this issue below). 
The fourth subset, \emph{Consecutive Live}, contains sessions that have at least two consecutive windows classified as live. Because of this consecutive occurrence, it is truly possible for the user to switch from VoD to live streaming. As shown in Table~\ref{tab:you_state} (under distribution column), \emph{Consecutive Live} sessions are far less frequent ($2\%$) than \emph{Stray Live} sessions ($6.4\%$).

The false-positive classifications (predominantly in Stray Live) interfere with our inferencing system, as once a session is classified as live, we begin measuring the experience of the session in terms of resolution bins and buffer stalls (\S\ref{sec:deploy_arch}). To tackle this problem, we introduce a state machine (Fig. 8) that helps us identify these false positives and make our system robust to misclassifications.

%



\begin{table*}[t]
	\vspace{-2mm}
	\begin{minipage}{0.48\linewidth}
		\centering
		\includegraphics[width=0.7\textwidth]{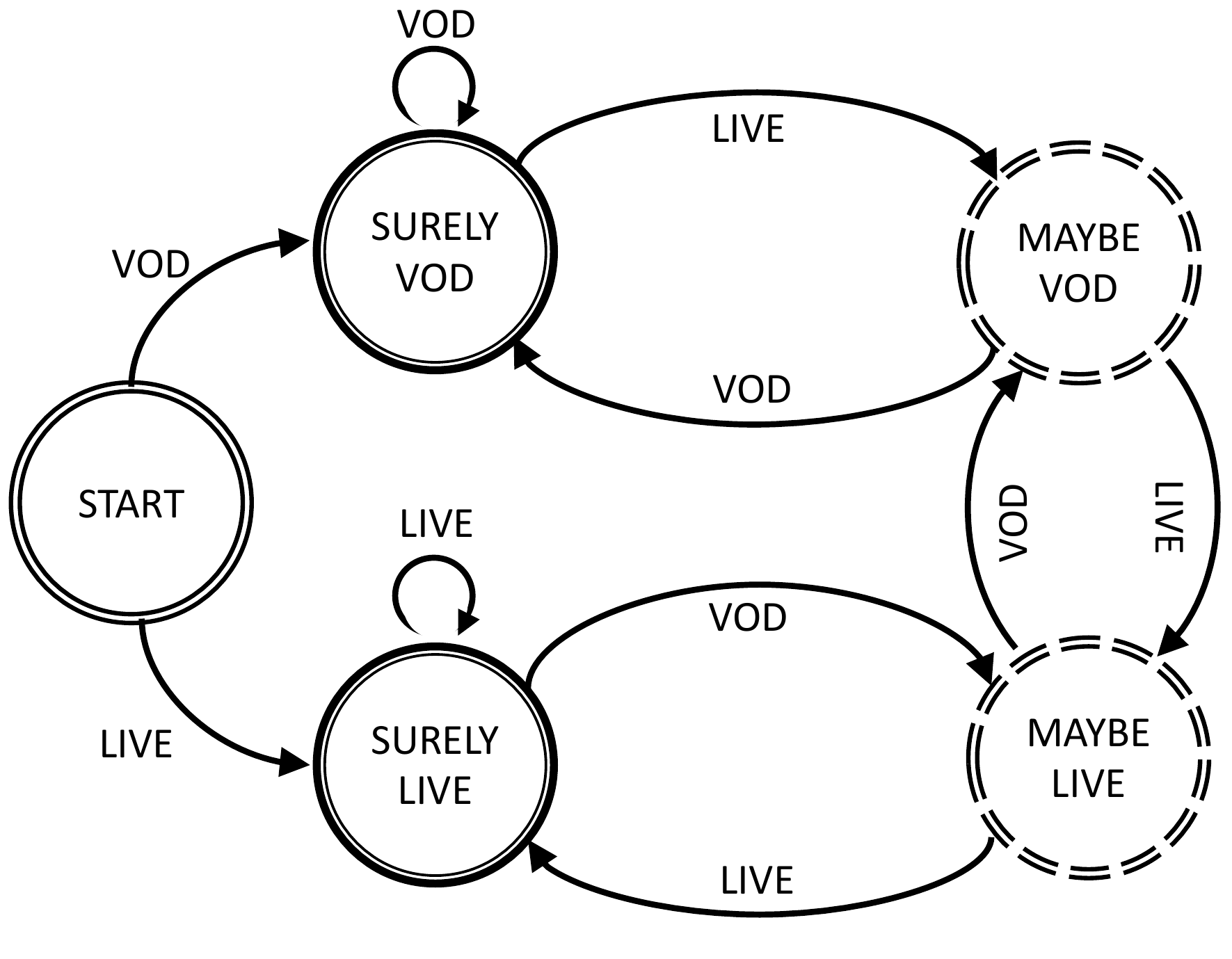}
		\vspace{-3mm}
		\captionof{figure}{Classification state machine.}
		\vspace{-3mm}
		\label{fig:state_machine}
	\end{minipage}
	\begin{minipage}{0.48\linewidth}
		\centering
		\caption{Aggregate QoE metrics of live videos: time fraction spent at various states.}
		\vspace{0mm}
		\label{tab:qoe_results}
		\begin{adjustbox}{max width=0.99\textwidth}	
			\begin{tabular}{@{}lrr@{}}
				\toprule
				\multicolumn{1}{l}{QoE states} & \multicolumn{1}{l}{Twitch} & \multicolumn{1}{l}{YouTube} \\ \midrule
				Stall  & 5.1\%  & 6.8\%  \\  
				Low-Def     & 3.9\%  & 19.1\% \\ 
				Standard-Def     & 30.8\% & 34.3\% \\ 
				High-Def     & 34.2\% & 46.6\% \\ 
				Source & 31.1\% & -      \\  \bottomrule
			\end{tabular}
		\end{adjustbox}
		
	\end{minipage}\hfill
	
\end{table*}

\begin{figure*}[t!]
	\begin{center}
		\mbox{
			\subfigure[Twitch.]{
				{\includegraphics[width=0.42\textwidth]{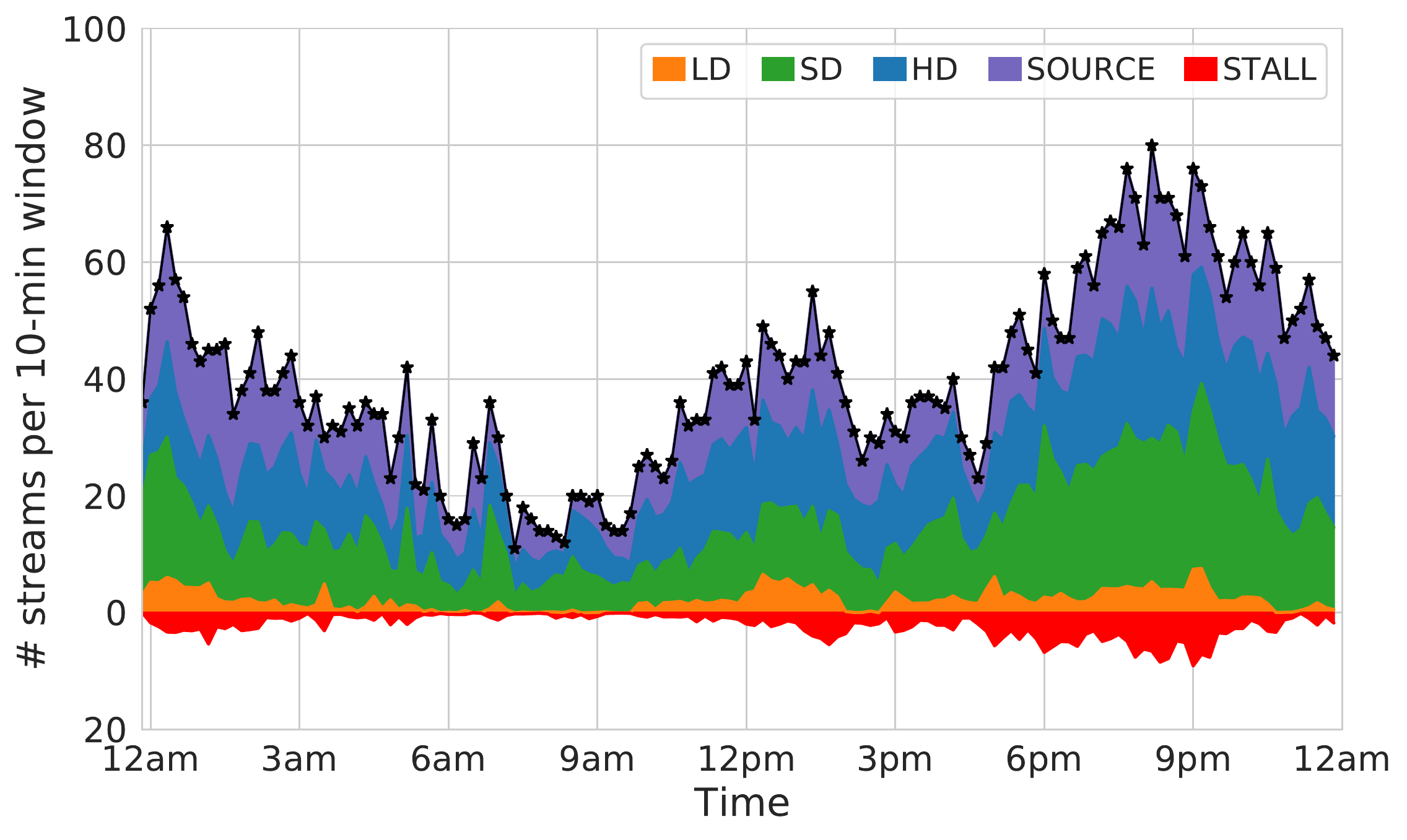}}\quad
				\label{fig:twqoe}
			}
		}
		\hspace{0mm}
		\mbox{
			\subfigure[YouTube.]{
				{\includegraphics[width=0.40\textwidth]{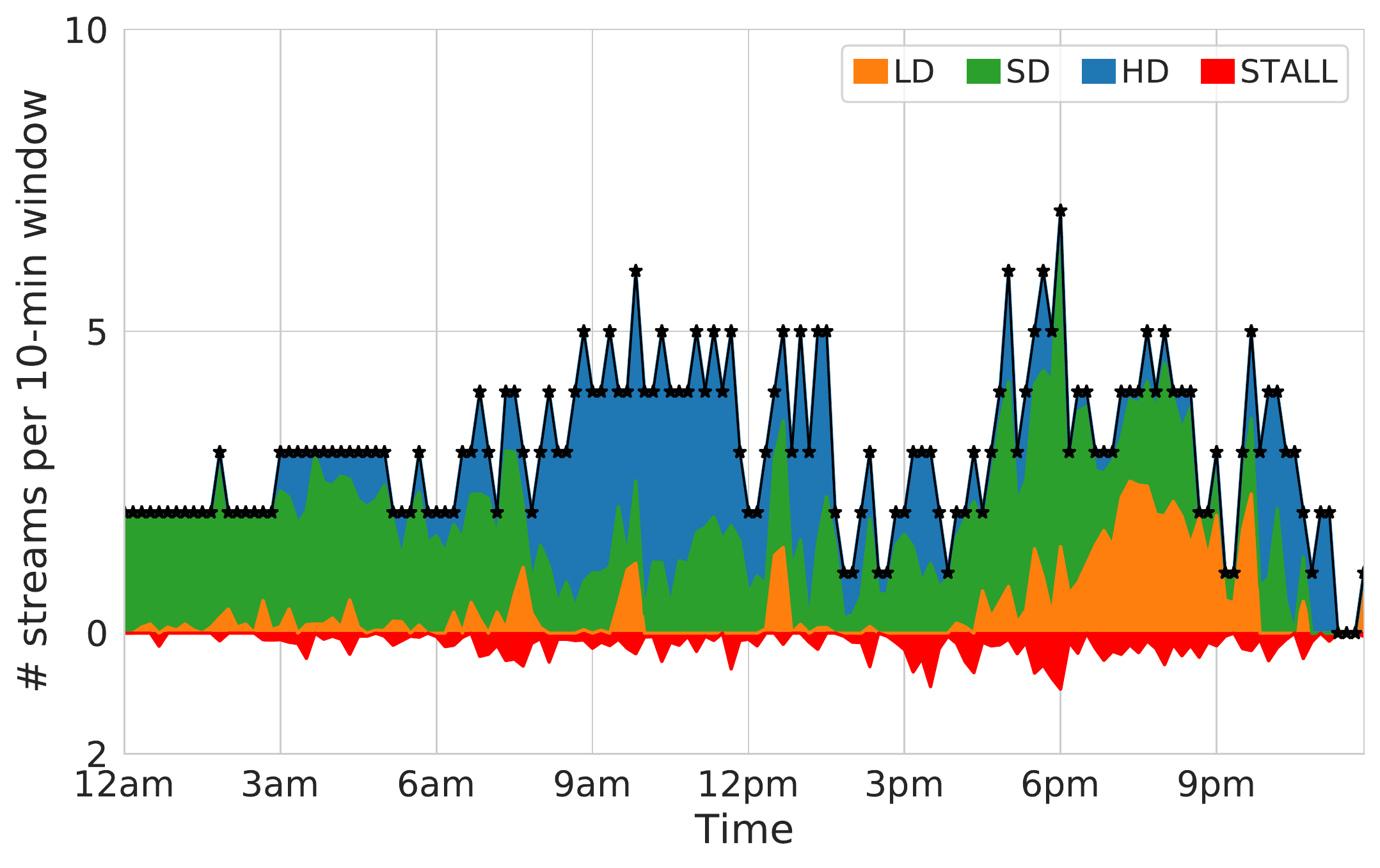}}\quad
				\label{fig:fbqoe}
			}
		}
		\vspace{-3mm}
		\caption{Daily trace of QoE predictions: (a) Twitch, and (b) YouTube. }
		\vspace{-4mm}
		\label{fig:QoEresults}
	\end{center}
\end{figure*}

The state machine shown in Fig.~\ref{fig:state_machine}, consists of five states including: (1) \emph{Start}, (2) \emph{Surely Live}, (3) \emph{Maybe Live}, (4) \emph{Surely VoD}, and (5) \emph{Maybe VoD}. A new session begins from the \emph{Start} state. It transitions to \emph{Surely VoD} or \emph{Surely Live} depending upon  whether the first window is classified VoD or live, respectively. 
Sessions in \emph{Only VoD} subset never transition out of the \emph{Surely VoD} state during its entire lifetime -- similarly, \emph{Only Live} sessions stay in the \emph{Surely Live} state (as show by self-loops in Fig.~\ref{fig:state_machine}).
The right-most column of Table~\ref{tab:you_state} reports for each category of YouTube sessions the total time fraction spent at each of the states. It is interesting to note that \emph{consecutive live} sessions spend almost $15$\% of the time in the \emph{Surely Live} state. 
Further analysis revealed that in many of these sessions, the user completely switches from VoD to Live, \ie the user was watching a VoD stream continuously for, say 10 minutes, and switched to Live streaming for 15 min thereafter. \emph{Stray Live} sessions, primarily stay at the state \emph{Surely VoD} (96.6\% of their lifetime), occasionally (2.4\% of the time) transition to the state \emph{Maybe VoD} (from \emph{Surely VoD}) due to a misclassified live window. 
Hence, we declare the prediction of a session (\textit{Live} or \textit{VoD}) only when it enters into one of the ``\emph{Surely}'' states, thus nullifying the impact of these false-positives. As a result, for a \emph{Stray Live} session, we move from \emph{Surely VoD} to \emph{Maybe Vod} and back to \emph{Surely VoD} without declaring the session as live. We also note that \emph{Stray Live} sessions end up in \emph{Surely Live} and \emph{Maybe Live} in very rare cases ($0.5$\% chance) due to a misclassification in the first window. These scenarios have little impact on our system since they are rare and eventually transition into \emph{Surely VoD}.

\subsection{QoE metrics in the Field}

Following classification, our system starts to report QoE metrics (resolution and buffer stalls) for the live video streams. As described in  \S\ref{sec:qoe}, our QoE models, fed by the network chunk metadata,  predict resolution and buffer value every 30 seconds. In what follows, we outline the results of analysis on raw data on an aggregate and daily basis. Note that YouTube QUIC has been excluded from our QoE study due to the challenges mentioned in \S\ref{sec:qoe}.

Table \ref{tab:qoe_results} shows a summary of live streaming QoE at the aggregate level. We observe that: (a) the ISP's network is performing relatively well, with more than $65$\% of  Twitch videos and over $45$\% of YouTube videos are played at HD quality and above, (b) while only $3.9$\% of Twitch videos were played at LD quality, a non-negligible fraction ($19.1$\%) of YouTube videos are played at LD quality, and (c) video stalls are experienced by only $5.1$\% of Twitch and $6.8$\% of YouTube live streams. Based on QoE statistics, one may infer that Twitch, a dedicated live streaming service, provides a better experience with higher resolution and fewer stalls.


In addition to the aggregate statistics, we plot in Fig.~\ref{fig:QoEresults} a daily trace of QoE predictions for both Twitch and YouTube on a representative weekday from our field trial period. The upper envelope black lines (with star markers) show the number of live streams per 10-min window. The shaded positive stack bars show the distribution of time fraction spent by all streams (during each 10-min window) in each of the resolutions. In contrast, the negative bar shows the average number of sessions facing a stall during each 10-min window.
It can be seen that the consumption of live streaming starts increasing (and peaks) at night hours. 
We believe that many live viewers (especially for Twitch) are gaming enthusiasts who often stay up late at night and watch content from worldwide. 
Another important observation is that QoE seems to be adversely affected for both Twitch and YouTube during evening peak hours \ie 7pm-10pm due to an increase in network activity (the total network load also peaks during the same period).

We believe further analysis can be carried out (beyond the scope of this work) on the collected metrics to gain insights like identifying users who continuously have poor QoE and/or abandon viewing after multiple resolution switches or buffer stall events. Such information would be beneficial to network operators in predicting support calls as well as churn rates. Our work demonstrates how real-time in-network identification and experience measurement of live video streaming can be realized. 

\subsection{Challenges and Limitations}\label{sec:challenges}
\textbf{Identifying video flows}: Our system primarily relies on the SNI field of the TLS handshake to determine whether a network flow carries video content, or not. During the course of this work, while we were able to classify TCP flows of all content providers (\ie is video or not), we have seen QUIC (used by YouTube) evolved from version 46 to version 50, wherein the SNI information is encrypted, thus no longer readable on the network. Further, with TLS 1.3 and eSNI \cite{chai2019importance}, this trend might soon be adopted by various content providers. In this work, for QUIC flows, we have used DNS information to tag a video server, and an empirical threshold of 4MB to identify elephant flows \cite{madanapalli2018real}, presumably carrying video content from that server. However, with the adoption of DoH and DNSSEC, even this information might not be available for a network operator, thus making it very challenging to identify video flows. In such a scenario, we may rely on behavioral traffic classification systems \cite{pacheco2018towards, iTeleScope19} to first identify video flows and then proceed to detect live video streams.
%

\textbf{Grouping flows into a video stream}: Some providers like YouTube use multiple flows to deliver video content of a session or re-use the same flows to play an entirely different video (if present on the same CDN server). 
This means that all flows from a client IP need to be combined, and then their aggregate features to be passed on to the classifier. In this work, we aggregate all video flows originating from a clientIP address into one session. This method has an inherent assumption that only one YouTube video is played at any point in time. While this assumption is more likely to be true for an IPv6 address (more than half of the traffic for YouTube was carried on IPv6) as each device has a unique IPv6 address, it might not always hold for IPv4 addresses. Multiple NAT-ted client devices which are concurrently watching YouTube content within a home network would share the same public IP address (v4). While automatically grouping flows into streams is beyond the scope of this work, we are aware that this is a limitation of our current system. We continue to develop algorithms in our future work that can do the grouping.
%
%

\textbf{Evolving live video streaming technologies}: Our system is able to identify live video streams from the production traffic that follow modern protocols such as HLS, LLHLS and formats such as CMAF. While we hope that the generality of the model (relying upon request counters) will still be able to distinguish the patterns, we are aware that a substantial change in streaming technology might require further work. We hope this work can serve as a stepping stone to build models for new live streaming technologies as the state of the Internet moves forward.
\section{Conclusion}
Live video streaming is a rapidly growing Internet application. ISPs today lack tools to infer QoE metrics of live streams in their network as existing DPI-based solutions fall short in detecting and monitoring live streams.
In this paper, we analyzed and released a dataset on playback metrics and network measurement of streams from Twitch and YouTube, identified key characteristics of live and on-demand streaming, and developed \textit{ReCLive}. This ML-based system distinguishes live streams from VoD streams using media requests patterns and infers QoE in terms of video resolution and buffer stall events for the detected live streams using chunk attributes extracted from the network flows.
\textit{ReCLive} works at scale and in real-time, as demonstrated by the field deployment in an ISP serving over 7000 home subscribers. By tracking the output of our classification models, we developed a state machine to make our live video detection mechanism more robust. 
Our method enables ISPs to better understand the experience of live streaming services purely using the behavioral profile of network flows, and subsequently enables them to take corrective actions to help improve user experience.

\bibliographystyle{IEEEtran}
\bibliography{livevideo}

\begin{IEEEbiography}[{\includegraphics[width=1in,height=1.25in,clip,keepaspectratio]{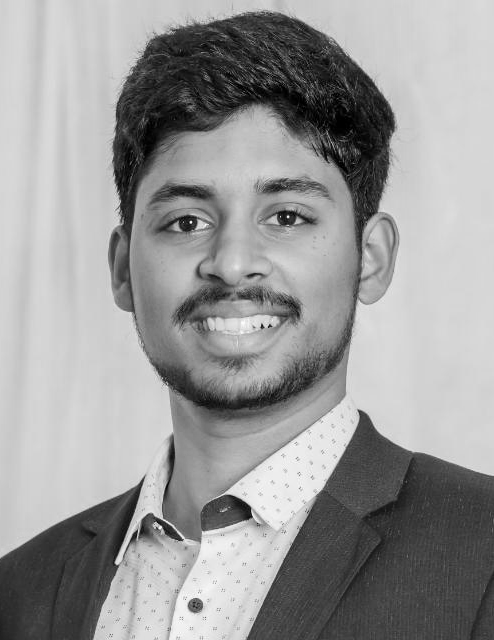}}]{Sharat~Chandra~Madanapali}
	received his B.E. degree in Computer Science from BITS Pilani in India. He is currently pursuing his Ph.D. in Electrical Engineering and Telecommunications at University of New South Wales (UNSW) in Sydney, Australia. His primary research interests include application QoE, applied machine learning and programmable networks.
\end{IEEEbiography}

\begin{IEEEbiography}[{\includegraphics[width=1in,height=1.35in,clip,keepaspectratio]{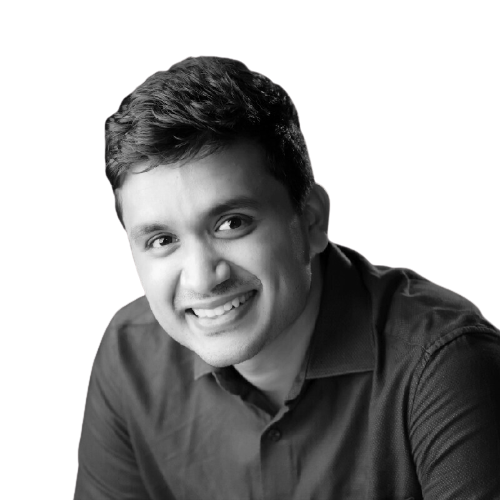}}]{Alex~Mathai}
	received his B.E. degree in Computer Science from BITS Pilani in India. He is currently a Research Software Engineer at IBM Research Labs, Bangalore. He is part of an effort to leverage AI tools and techniques to accelerate adoption of the hybrid cloud paradigm. His primary research interests include data driven modeling and applied deep learning.
\end{IEEEbiography}

\begin{IEEEbiography}[{\includegraphics[width=1in,height=1.25in,clip,keepaspectratio]{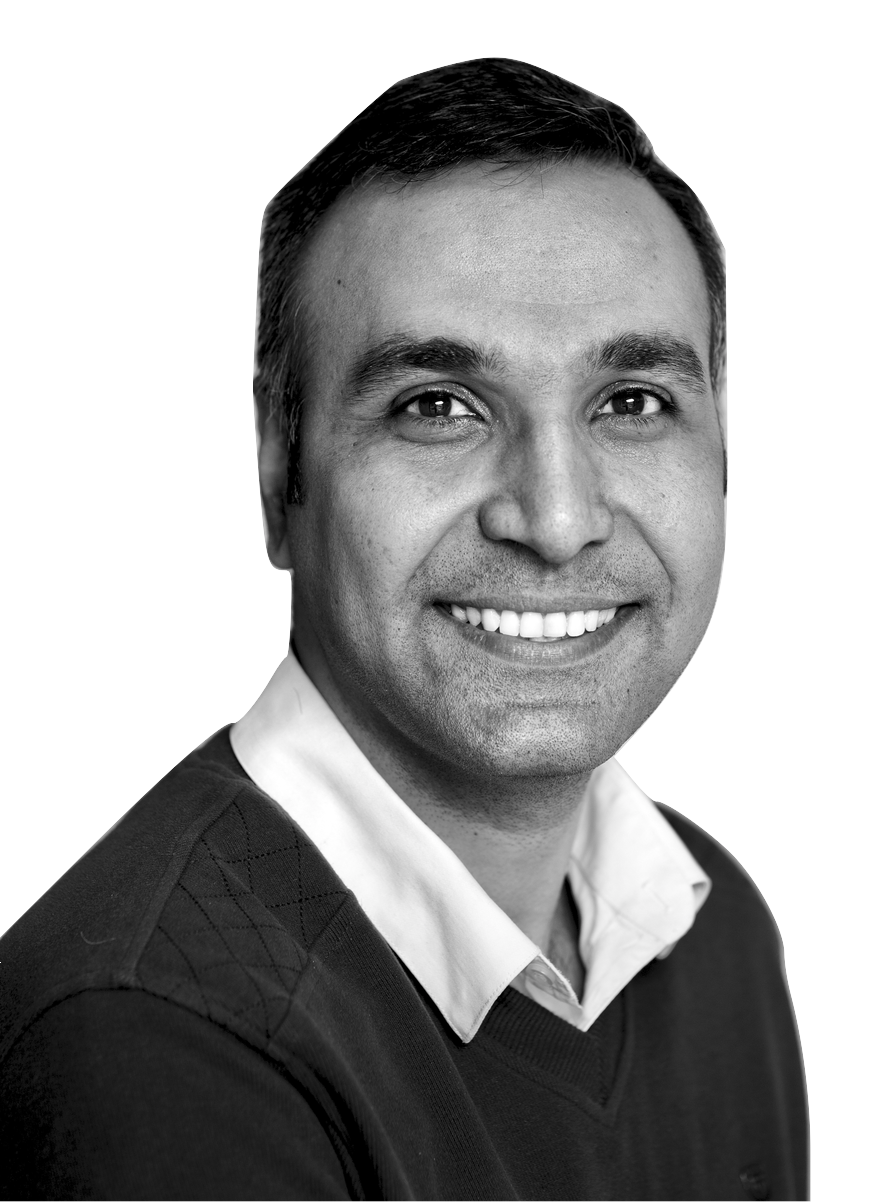}}]{Hassan Habibi Gharakheili}
	received his B.Sc. and M.Sc. degrees of Electrical Engineering from the Sharif University of Technology in Tehran, Iran in 2001 and 2004 respectively, and his Ph.D. in Electrical Engineering and Telecommunications from the University of New South Wales (UNSW) in Sydney, Australia in 2015. He is currently a Senior Lecturer at UNSW Sydney. His research interests include programmable networks, learning-based networked systems, and data analytics in computer systems.
\end{IEEEbiography}

\begin{IEEEbiography}[{\includegraphics[width=1in,height=1.25in,clip,keepaspectratio]{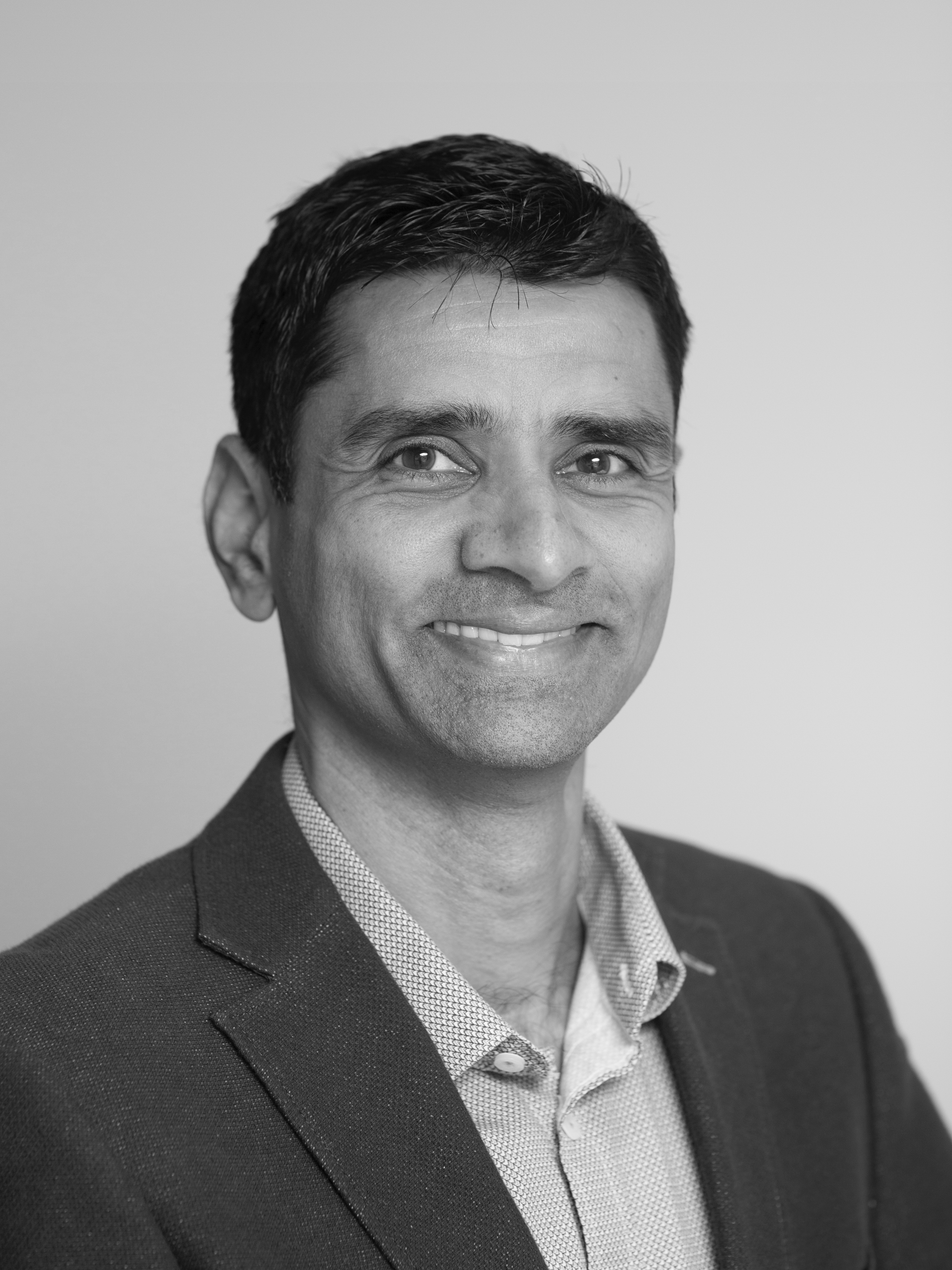}}]{Vijay Sivaraman}
	received his B. Tech. from the Indian Institute of Technology in Delhi, India, in 1994, his M.S. from North Carolina State University in 1996, and his Ph.D. from the University of California at Los Angeles in 2000. He has worked at Bell-Labs as a student Fellow, in a silicon valley start-up manufacturing optical switch-routers, and as a Senior Research Engineer at the CSIRO in Australia. He is now a Professor at the University of New South Wales in Sydney, Australia. His research interests include Software Defined Networking, network architectures, and cyber-security particularly for IoT networks.
\end{IEEEbiography}	

\balance

\end{document}